# GRB 080319B: A Naked-Eye Stellar Blast from the Distant Universe

J. L. Racusin[1], S.V. Karpov[2], M. Sokolowski[3], J. Granot[4], X. F. Wu[1,5], V. Pal'shin[6], S. Covino[7], A.J. van der Horst[8], S. R. Oates[9], P. Schady[9], R. J. Smith[10], J. Cummings[11], R.L.C. Starling[12], L. W. Piotrowski[13], B. Zhang[14], P.A. Evans[12], S. T. Holland[15,16,17], K. Malek[18], M. T. Page[9], L. Vetere[1], R. Margutti[19], C. Guidorzi[7,10], A. Kamble[20], P.A. Curran[20], A. Beardmore[12], C. Kouveliotou[21], L. Mankiewicz[18], A. Melandri[10], P.T. O'Brien[12], K.L. Page[12], T. Piran[22], N. R. Tanvir[12], G. Wrochna[3], R.L. Aptekar[6], C. Bartolini[28], S. Barthelmy[11], G. M. Beskin[2], S. Bondar[23], S. Campana[7], A. Cucchiara[1], M. Cwiok[13], P. D'Avanzo[7], V. D'Elia[24], M. Della Valle[25,26], W. Dominik[13], A. Falcone[1], F. Fiore[24], D. B. Fox[1], D. D. Frederiks[6], A. S. Fruchter[27], D. Fugazza[7], M. Garrett[28,29,30], N. Gehrels[11], S. Golenetskii[6], A. Gomboc[31], G. Greco[32], A. Guarnieri[32], S. Immler[15,17], G. Kasprowicz[33], A. J. Levan[34], E.P. Mazets[6], E. Molinari[7], A. Moretti[7], K. Nawrocki[3], P.P. Oleynik[6], J. P. Osborne[12], C. Pagani[1], Z. Paragi[35], M. Perri[36], A. Piccioni[32], E. Ramirez-Ruiz[37], P. W. A. Roming[1], I. A. Steele[10], R. G. Strom[20,28], V. Testa[24], G. Tosti[37], M.V. Ulanov[6], K. Wiersema[12], R. A. M. J. Wijers[20], A. F. Zarnecki[13], F. Zerbi[7], P. Mészáros[1,38], G. Chincarini[7,19], D. N. Burrows[1]

[1]Department of Astronomy & Astrophysics, 525 Davey Lab, Pennsylvania State University, University Park, PA, USA
[2]Special Astrophysical Observatory, Nizhnij Arkhyz, Zelenchukskij region, Karachai-Cirkassian Republic, Russia 369167
[3]Soltan Institute for Nuclear Studies, 05-400 Otwock-Swierk, Poland
[4]Centre for Astrophysics Research, University of Hertfordshire, College Lane, Hatfield, Herts, AL10 9AB, UK
[5]Purple Mountain Observatory, Chinese Academy of Sciences, Nanjing 210008, China
[6]Ioffe Physico-Technical Institute, Laboratory for Experimental Astrophysics, Saint Petersburg 194021, Russian Federation
[7]INAF-Osservatorio Astronomico di Brera, via E. Bianchi 46, I-23807 Merate (LC), Italy
[8]NASA Postdoctoral Program Fellow, NSSTC, 320 Sparkman Drive, Huntsville, AL 35805, USA
[9]The UCL Mullard Space Science Laboratory, Holmbury St. Mary, Surrey, RH5 6NT, UK
[10]Astrophysics Research Institute, Liverpool John Moores University, Twelve Quays House, Birkenhead, CH41 1LD, UK
[11]Astrophysics Science Division, Code 661, NASA's Goddard Space Flight Centre, 8800 Greenbelt Rd, Greenbelt, MD 20771, USA
[12]Department of Physics & Astronomy, University of Leicester, Leicester, LE1 7RH, UK
[13]Institute of Experimental Physics, University of Warsaw, Hoża 69, 00-681 Warsaw, Poland
[14]Department of Physics and Astronomy, University of Nevada Las Vegas, Las Vegas, NV 89154, USA
[15]Astrophysics Science Division, Code 660.1, NASA's Goddard Space Flight Centre, 8800 Greenbelt Rd, Greenbelt, MD 20771, USA
[16]Universities Space Research Association, 10211 Wincopin Circle, Suite 500, Columbia, MD 21044, USA
[17]Centre for Research and Exploration in Space Science and Technology, Code 668.8, NASA's Goddard Space Flight Centre, 8800 Greenbelt Road, Greenbelt, MD 20771, USA
[18]Center for Theoretical Physics PAS, Al. Lotników 32/46, 02-668 Warsaw, Poland
[19]Universita` degli Studi di Milano Bicocca, Physics Dept., Piazza della Scienza 3, I-20126 Milano, Italy
[20]Astronomical Institute "Anton Pannekoek", University of Amsterdam, Kruislaan 403, 1098 SJ Amsterdam, The Netherlands
[21]NASA/Marshall Space Flight Center, VP62, NSSTC, 320 Sparkman Drive, Huntsville Al 35805, USA
[22]Racah Institute for Physics, The Hebrew University, Jerusalem, 91904, Israel
[23]Institute for Precise Instrumentation, Nizhnij Arkhyz, Russia, 369167
[24]INAF - Osservatorio Astronomico di Roma, via Frascati 33, 00040 Monteporzio Catone, Italy

[25]European Southern Observatory, Karl-Schwarzschild-Strasse 2D-85748 Garching bei München
[26]INAF-Osservatorio Astronomico di Capodimonte, Salita Moiariello, 1680131, Napoli, Italy
[27]Space Telescope Science Institute, 3700 San Martin Drive, Baltimore, MD 21218, USA
[28]Netherlands Institute for Radio Astronomy (ASTRON), Postbus 2, 7990 AA Dwingeloo, The Netherlands
[29]Leiden Observatory, University of Leiden, P.B. 9513, Leiden 2300 RA, the Netherlands
[30]Centre for Astrophysics and Supercomputing, Swinburne University of Technology, Hawthorn, Victoria 3122, Australia.
[31]Faculty of Mathematics and Physics, University of Ljubljana, Jadranska 19, SI-1000 Ljubljana, Slovenia
[32]Universitá di Bologna, Via Ranzani 1 - 40126, Bologna, Italy
[33]Institute of Electronic Systems, Warsaw University of Technology, Nowowiejska 15/19, 00-665 Warsaw, Poland
[34]Department of Physics, University of Warwick, Coventry CV4 7AL
[35]Joint Institute for VLBI in Europe (JIVE), Netherlands Institute for Radio Astronomy (ASTRON), Postbus 2, 7990 AA Dwingeloo, The Netherlands
[36]ASI Science Data Center, c/o ESRIN, via G. Galilei, I-00044 Frascati, Italy
[36]Department of Astronomy and Astrophysics, University of California, Santa Cruz, CA 95064, USA
[37]University of Perugia, Piazza dell'Università, 1 - 06100, Perugia, Italy
[38]Physics Department, 104 Davey Lab, Pennsylvania State University, University Park, PA 16801, USA

**Long duration gamma-ray bursts (GRBs) release copious amounts of energy across the entire electromagnetic spectrum, and so provide a window into the process of black hole formation from the collapse of a massive star. Over the last forty years, our understanding of the GRB phenomenon has progressed dramatically; nevertheless, fortuitous circumstances occasionally arise that provide access to a regime not yet probed. GRB 080319B presented such an opportunity, with extraordinarily bright prompt optical emission that peaked at a visual magnitude of 5.3, making it briefly visible with the naked eye. It was captured in exquisite detail by wide-field telescopes, imaging the burst location from before the time of the explosion. The combination of these unique optical data with simultaneous γ-ray observations provides powerful diagnostics of the detailed physics of this explosion within seconds of its formation. Here we show that the prompt optical and γ-ray emissions from this event likely arise from different spectral components within the same physical region located at a large distance from the source, implying an extremely relativistic outflow. The chromatic behaviour of the broadband afterglow is consistent with viewing the GRB down the very narrow inner core of a two-component jet that is expanding into a wind-like environment consistent with the massive star origin of long GRBs. These circumstances can explain the extreme properties of this GRB.**

The exceptional GRB 080319B, discovered by NASA's *Swift* GRB Explorer mission[1] on 19 March 2008, set new records among these most luminous transient events in the Universe. GRBs are widely thought to occur through the ejection of a highly relativistic, collimated outflow (jet), produced by a newly formed black hole. Under the standard fireball model[2,3,4,5,6], collimated relativistic shells propagate away from the central engine, crash into each other (internal shocks), and decelerate as they plough into the surrounding medium (external/forward shocks). Reverse shocks propagate back into the jet, generating optical emission. With a uniquely bright peak visual magnitude of 5.3 (Figure 1) at a redshift of $z=0.937$ (ref. 7), GRB 080319B was the most luminous optical burst ever observed. During the first 40 seconds of the event, an observer in a dark location could have seen the prompt optical emission from the source with the naked eye. The astronomical community has been waiting for such an event for the last nine years, ever since GRB 990123 (the previous record holder for the highest peak optical brightness) peaked at a visual magnitude of ~9, leading to significant insight into the GRB optical emission mechanisms[8]. The location of GRB 080319B was fortuitously only 10° away from another event, GRB 080319A, also detected by *Swift* less than 30 minutes earlier, allowing several wide field telescopes to detect the optical counterpart of GRB 080319B instantly. The rapid localization by *Swift* enabled prompt multi-wavelength follow-up observations by robotic ground-based telescopes, resulting in arguably the best broadband GRB observations ever obtained. These observations continued for weeks afterwards as we followed the fading afterglow, providing strong constraints on the physics of the explosion and its aftermath.

At its peak, GRB 080319B displayed the brightest optical and X-ray fluxes ever measured for a GRB, and one of the highest γ-ray fluences recorded. Previous early optical observations of GRBs lacked both the temporal resolution to probe the optical flash in detail, and the accuracy needed to trace the transition from the prompt emission within the outflow to external (reverse and forward) shocks caused by interaction with the progenitor environment. Our broadband data cover 11.5 orders of magnitude in wavelength, from radio to γ-rays, and begin (in the optical and γ-ray bands) *before* the explosion. We can for the first time identify three different components responsible for the optical emission. The earliest data (at $t \equiv T-T_0 < 50$ s) provide evidence that the bright optical and γ-ray emissions stem from the same physical region within the outflow. The second optical component (50 s $< t <$ 800 s) shows the distinct characteristics of a reverse shock, while the final component (at $t > 800$ s) represents the afterglow produced as the external forward shock propagates into the surrounding medium. Previous measurements of GRBs have revealed only one or two of these components at a time[9,10,11], but never all three in the same burst with such clarity. GRB 080319B is, therefore, a test-bed for broad theoretical modelling of GRBs and their environments.

**GRB 080319B: Discovery and Multi-wavelength Observations**

The *Swift*-Burst Alert Telescope (BAT[12]; 15-350 keV) triggered on GRB 080319B at[13] $T_0$ = 06:12:49 UT on March 19, 2008. The burst direction was already within the field of view of the BAT for 1080 s prior to its onset (with a data gap from $T_0$-965 to $T_0$-900 s), placing strong limits on any precursor emission. The burst was simultaneously detected with the Konus γ-ray detector onboard the Wind satellite[14,15], which is sensitive to energies between 20 keV and 15 MeV. Both BAT and Konus-Wind (KW) light curves (Supplementary Figures 1 and 3) show a complex, strongly energy-dependent structure, with many clearly separated pulses above 70 keV and a generally smoother behaviour at lower energies, lasting approximately 57 s.

The time-averaged KW γ-ray spectrum is well fit using a Band function[16], with a low-energy slope of $a = -0.855^{+0.014}_{-0.013}$, a high-energy slope of $b = -3.59^{+0.32}_{-0.62}$, and peak energy of $E_p$=675±22 keV ($\chi^2/dof$ = 110.4/80). Time-resolved KW spectra show that the Band function parameters vary rapidly during the prompt emission, with the low energy slope changing from -0.5 to -0.9 and $E_p$ changing from ~740 keV to ~540 keV in the first 30 s (see Supplementary Table 1 and Supplementary Figure 3). Time-resolved single power-law spectral fits of the BAT data show the photon index shifting rapidly from ~ 1.0 to ~ 2.1 at $T_0$+53s (near the end of the prompt phase; Supplementary Figure 2). The burst had a peak flux of $F_p$=2.26 ± 0.21 x$10^{-5}$ erg cm$^{-2}$ s$^{-1}$, fluence of $F_\gamma$ (20 keV – 7 MeV) = (6.13 ± 0.13)×$10^{-4}$ erg cm$^{-2}$, peak isotropic equivalent luminosity of $L_{p,iso}$=1.01±0.09 ×10$^{53}$ erg s$^{-1}$ (at the luminosity distance of $d_L$=1.9×10$^{28}$ cm assuming cosmological parameters $H_0$ = 71 km s$^{-1}$ Mpc$^{-1}$, $\Omega_M$ = 0.27, $\Omega_\Lambda$ = 0.73), and the isotropic equivalent γ-ray energy release of $E_{\gamma,iso}$= 1.3×10$^{54}$ erg (20 keV – 7 MeV). These are among the highest ever measured. All quoted errors are 90% confidence throughout this paper, unless specified otherwise.

The wide-field robotic optical telescope "Pi of the Sky"[17,18] (located at Las Campanas Observatory), and the wide-field robotic instrument Telescopio Ottimizzato per la Ricerca dei Transienti Ottici RApidi (TORTORA[19], which is attached to the 60 cm robotic optical/near-infrared Rapid Eye Mount [REM[20]] telescope located at La Silla), both coincidentally had the GRB within their fields of view at the time of the explosion (as they were both already observing GRB 080319A (ref .21)). "Pi of the Sky" observed the bright optical transient, which began at 2.75 ± 5 s after the BAT trigger, rose rapidly, and peaked at ~$T_0$+18 s; the telescope observed the transient until it faded below threshold to ~12$^{th}$ magnitude after 5 minutes[22]. TORTORA measured the brightest portion of the optical flash with high time resolution, catching 3 separate peaks (Figure 1), enabling us to do detailed comparisons between the prompt optical and γ-ray emission.

The *Swift* spacecraft and the REM telescope both initiated automatic slews to the burst, resulting in optical observations in the R band (REM) and white light (1700-6000 Å, with the *Swift* UltraViolet-Optical Telescope, UVOT[23]) beginning at $T_0$+51 s and 68 s, respectively. The *Swift* X-ray Telescope (XRT[24]) began observing the burst at $T_0$+51 s, providing time-resolved spectroscopy in the 0.3-10 keV band. Over the next several hours, we obtained ultraviolet, optical and near-infrared (NIR) photometric observations of the GRB afterglow with the *Swift*-UVOT, REM, the Liverpool Telescope, the Faulkes Telescope North, Gemini-North, and the Very Large Telescope (VLT). Subsequent optical spectroscopy by Gemini-N and the Hobby-Eberly Telescope (HET) confirmed the redshift of 0.937 (Supplementary Figures 4 and 5). X-ray and optical observations continued for more than four weeks after the burst. Multiple epochs of radio observations with the Westerbork Synthesis Radio Telescope (WSRT) revealed a radio counterpart ~2-3 days after the burst. The composite broadband light curves of GRB080319B, which include all data discussed throughout this paper, and which cover 8 orders of magnitude in flux and over 6 orders of magnitude in time, are shown in Figure 2 and summarized in Table 1.

**Prompt Emission from an Ultra-Relativistic Outflow**

The temporal coincidence of the bright "optical flash" and the γ-ray burst (Figure 1) provides important constraints on the nature of the prompt GRB emission mechanism. While there is a general consensus that the prompt γ-rays must arise from internal dissipation within the outflow, likely due to internal shocks, the optical flash may either arise from the same emitting region as the γ-rays, or from the reverse shock that decelerates the outflow as it sweeps-up the external medium. The reverse shock becomes important when the inertia of the swept-up external matter starts to slow down the ejecta appreciably, at a larger radius than the internal shocks dissipation.

The temporal coincidence of the onset and overall shape of the prompt optical and γ-ray emissions suggest that both originate from the same physical region (see also ref. 25), though their respective peaks during this phase do not correlate (see Supplementary Figures 8 and 9 and the related discussion in the Supplementary Materials). Nevertheless, the initial steep rise (at $t < 18$ s) and the rapid decline (at $t > 43$ s) indicate[26,27] that the optical flash did not arise from a reverse shock (cf. GRB 990123, ref. 28,29). Also, the optical pulse widths are expected to grow at least linearly in time in this model[30], yet they remain unchanged. The apparent larger variability time and possible slight temporal lag relative to the γ-rays may be the result of the optical being somewhat below the synchrotron self-absorption frequency, $\nu_a$, which results in wider pulses that peak later compared to the γ-rays in the internal shocks model.

The flux density of the optical flash is ~$10^4$ times larger than the extrapolation of the γ-ray spectrum into the optical band (Figure 3). The popular interpretation of the soft γ-rays as synchrotron emission cannot account for such a distinct lower energy spectral component from the same physical region, suggesting that different radiation mechanisms must dominate in each regime. The most natural (but by no means the only viable) candidates are synchrotron for the optical and synchrotron self-Compton (SSC) for the γ-rays[31,32]. The Compton $Y$ parameter, defined as the ratio of the inverse Compton to synchrotron energy losses, is $Y \sim \nu F_\nu(E_p)/\nu F_\nu(E_{p,syn}) \gtrsim 10$, where $E_{p,syn}$ is the peak photon energy of the synchrotron $\nu F_\nu$ spectrum, to account for the fact that the prompt γ-ray energy is higher than the prompt optical/UV synchrotron energy. This would imply a third spectral component arising from second-order inverse-Compton scattering that peaks at energies around $E_{p,2} \approx E_p^2/E_{p,syn} \approx 23(E_{p,syn}/20 \text{ eV})^{-1}$ GeV. Note that the Klein-Nishina suppression becomes important only at $E > 94(E_{p,syn}/20 \text{ eV})^{-1/2} \Gamma_3$ GeV, where $\Gamma = 10^3 \Gamma_3$ is the outflow bulk Lorentz factor. This third spectral component carries more energy than the observed γ-rays, by a factor $Y \gtrsim 10$, changing the energy budget of this burst and implying that GRB 080319B was even more powerful than inferred from the observed emission. Most of the energy in this burst was emitted in this undetected GeV component, which would have been detected by the AGILE satellite had it not been occulted by the Earth, and would have been easily detectable by the upcoming *GLAST*[33] satellite.

Such bright prompt optical flashes are rare. The exceptional brightness of the optical flash in GRB 080319B implies that $v_a$ cannot be far above the optical band near the peak time. The optical brightness temperature implies that $300 \leq \Gamma(t_v/3 \text{ s})^{2/3} \leq 1400$, and therefore $\Gamma \sim 10^3$, where $t_v \equiv R\Gamma^{-2}c^{-1}$ is the rough variability timescale in the internal shocks model. Because of the extremely high bulk Lorentz factor $\Gamma$, the internal shocks occur at an unusually large radius $R = 10^{16} R_{16}$ cm, where $0.8 \leq R_{16} (t_v/3\text{s})^{1/3} \leq 20$, resulting in a relatively low self-absorption frequency $v_a$, which in turn allows the optical photons to escape.

Since the observed γ-ray emission of GRB 080319B shows very similar properties to those of most GRBs, it may be representative of the main underlying physical mechanism. If so, similar lower energy spectral components would be expected in most GRBs. The paucity of bright optical flashes may be attributed to less relativistic outflows in most GRBs, leading to smaller emitting radii $R$, higher optical depths, and significantly higher values of $v_a$, ultimately suppressing the optical emission. In this picture, the spectacular optical brightness of GRB 080318B is mainly due to its unusually large $\Gamma$.

**Empirical Fits to the Broadband Afterglow Data**

Our broadband dataset enabled us to measure the temporal and spectral evolution of GRB 080319B throughout the afterglow. The radiation mechanism is assumed to be synchrotron emission, as postulated in the standard fireball model[2,3,4,5,6], and the spectral and light curve segments are fit with power-laws, described by $F_v = t^{-\alpha} v^{-\beta}$, with decay index $\alpha$ and spectral energy index $\beta$.

After the optical flash, the optical light curve is best described by the superposition of three different power-law components (Supplementary Figure 6), with decay indices of $\alpha_{opt,1}$ = 6.5±0.9 (the tail of the optical flash), $\alpha_{opt,2}$ = 2.49±0.09, and $\alpha_{opt,3}$ = 1.25±0.02. The X-ray light curve clearly differs from the optical light curve during the first ~12 hours (Fig. 2). After a short (~30 s) flat smooth transition from the tail of the γ-ray prompt emission, the X-ray light curve after ~80 s is best fit by a triple broken power-law with decay indices of 1.44±0.07, 1.85±0.10, $1.17^{+0.14}_{-0.23}$, and $2.61^{+2.04}_{-0.91}$, and with breaks times of 2242±940 s, $4.1^{+2.8}_{-1.7} \times 10^4$ s, and 1.0±0.5×10$^6$ s ($\chi^2/dof$=880/697=1.26).

We created broadband spectral energy distributions (SEDs) at 11 epochs ranging from $T_0$+150 s to $T_0$+500 ks. The SEDs can be fit by power-laws, broken power-laws, or double broken power-laws. Deviations from broken power-laws at the UV and soft X-ray frequencies allow us to measure the host galaxy extinction and X-ray absorption column density (see discussion in Supplementary Materials). Our resulting best-fit spectral models are described in detail in the Supplementary Materials and demonstrated in Supplementary Figures 10, 11, and 12.

**Interpretation of the Chromatic Afterglow**

Following the prompt phase, the early (minutes to hours) X-ray and optical behaviour are inconsistent with the predictions of the standard afterglow theory, suggesting that they must stem from different emission regions. In particular, we find that the optical, X-ray, and γ-ray emission from this burst are explained reasonably well by a two-component jet model[34,35,36,37,38,39] (Figure 4, Table 2), consisting of an ultra-relativistic narrow jet with a jet break time of $t_{b,1}$ ~ 2800 s, surrounded by a broader, less energetic jet with a lower Lorentz factor and a jet break time of $t_{b,2}$ ~ 1 Ms. The empirical triple broken power-law of the X-ray light curve is then interpreted as the superposition of two broken power-law components representing these two jets (Supplementary Figure 7), with slopes of $\alpha_{x,1}=1.45\pm0.05$, $\alpha_{x,2}= 2.05^{+0.44}_{-0.27}$ for the narrow jet, and $\alpha_{x,3}=0.95^{+0.19}_{-0.69}$, $\alpha_{x,4}=2.70^{+2.06}_{-1.12}$ for the wide jet ($\chi^2$/dof = 903/700 = 1.29). This structure, where the Lorentz factor and energy per solid angle are highest near the axis and decreases outwards, either smoothly or in quasi-steps qualitatively resembles the results of numerical simulations of jet formation in collapsars[40]. Further details of the model are given in the Supplementary Materials; here we summarize the model results and apply them to the observational data.

The optical light curve between 50 s < $t$ < 800 s is dominated by the second optical power-law component, which we interpret as emission from the ***reverse shock*** associated with the interaction of the ***wide jet*** with the external medium. This segment has $\alpha_{opt,2}$ = 2.49±0.09 and $\beta_{opt,2}$ = 0.49±0.14, consistent with the expectations for the high-latitude emission[41] from a reverse shock ($\alpha=2+\beta$) if the cooling frequency, $\nu_c$, is below the optical band and the injection frequency $\nu_m$ > $10^{16}$ Hz. Emission from the reverse shock peaks around $t$~50 s in the optical with a peak flux density of ~2-3 Jy, but is initially overwhelmed by the much brighter prompt emission and does not become visible until the latter dies away. The high peak luminosity of the optical reverse shock component soon after the end of the γ-ray emission indicates that the reverse shock was at least mildly relativistic. The GRB outflow could not have been highly magnetized ($\sigma \gg 1$) when it crossed the reverse shock, or the reverse shock would have been suppressed[42], implying $\sigma \lesssim 1$, where $\sigma$ is the electromagnetic to kinetic energy flux ratio. Although $\sigma \ll 1$ allows a strong reverse shock, it suppresses its optical emission. Therefore, $\sigma \sim 0.1$–1 is needed to obtain the observed bright emission from the reverse shock[43,44].

On the other hand, the X-ray light curve in the interval 50 s < $t$ < 40 ks is dominated by the ***forward shock*** of the ***narrow jet*** component interacting with a surrounding medium produced by the wind[45] of the progenitor star in the slow cooling case ($\nu_m < \nu_x < \nu_c$). The first break in the X-ray light curve at $t_{b,1}$=2808 ± 913 s is attributed to a jet break[46] in this narrow jet, leading to a jet opening half-angle of $\theta_j \sim$ 0.2°. The beaming-corrected jet energy for the narrow jet is then $E_j = E_{k,iso}\, \theta_j^2/2 =$ 2.1×10$^{50}$ erg. Since this break is not seen in the optical light curve, the optical flux from the narrow jet forward shock must be much less than that of the wide jet, implying $\nu_{opt} < \nu_m < \nu_x < \nu_c$.

The optical emission after $T_0 + 800$ s is dominated by a single power-law with $\alpha_{opt,3} = 1.25\pm0.02$ and $\beta_{opt,3} = 0.50\pm0.07$, consistent with the expectation for ***forward shock*** emission from the ***wide jet*** with $\nu_m < \nu_{opt} < \nu_c$. The late X-ray afterglow after 40 ks is also dominated by the forward shock of the wide jet with an overall spectrum of $\nu_m < \nu_{opt} < \nu_c < \nu_x$. At approximately 11 days post-burst, the X-ray light curve breaks to a steeper slope (confirmed by a late *Chandra* observation, E. Rol, private communication). If this break is interpreted as the jet break of the wide jet, it corresponds to an initial jet half-opening angle of ~4°, which in turn implies a beaming-corrected energy for the wide jet of $E_j = 1.9\times10^{50}$ erg. The wide jet forward shock also accounts for the observed radio emission, which is strongly modulated by the effects of Galactic scintillation (see Methods section for more detailed discussion)[47,48] when the source is small.

**Discussion**

We have shown that the extraordinary optical brightness of this burst is due to an unusually large bulk Lorentz factor, in addition to its overall very large isotropic equivalent energy, distinguishing this event from typical GRBs. The optical and γ-ray prompt emissions stem from the same physical region, yet are attributed to different spectral emission mechanisms, likely synchrotron and SSC, respectively.

With an appropriate choice of parameters, the afterglow of GRB 080319B can be well described by a two-component jet model, with a very narrow (~0.2°) and highly relativistic jet, coaxial with a wider (4°) jet having more conventional properties. This interpretation requires a few caveats: both jet breaks are sharper than one would expect for a jet propagating into a wind environment, the SED models only qualitatively represent the data and require slight deviations in the scaling laws to obtain good fits, and coincidences are needed to explain the bright prompt emission with the narrow jet pointed at us. Considering the rarity of this event, these coincidences are probably statistically acceptable. Nevertheless, alternative models such as a blast-wave propagating into a complex medium (see Supplemental Figures 13 and 14, and related discussion in the Supplemental Materials), or evolving microphysical parameters, may also fit at least some aspects of the data, but we consider the two-component jet model to be most plausible interpretation. An interesting consequence of these theoretical considerations is that GRB 080319B, which has the best broadband dataset ever recorded, is not consistent with the expectations of any of the simple GRB models previously studied. The case for multiple spectral emission components and the two-component jet presented here suggests that similar models may be able to explain at least some of the chromatic breaks seen in optical and X-ray afterglows over the last few years that have been difficult to reconcile with the standard models[3,4].

GRB 080319B is unparalleled in terms of optical and X-ray flux, γ-ray fluence, multi-wavelength broadband long temporal coverage, and energetics. The circumstances that led to the exceptional properties of this event were largely serendipitous, yet enlightening. We deduce that we happened to view this monster down the barrel of the very narrow and energetic jet. The probability of observing within this tiny solid angle is small ($\sim 10^{-3}$). If every GRB has such a narrow jet, we should expect to detect the narrow jet emission from a GRB every ~3-10 years. Had we observed GRB 080319B even slightly off-axis, the behavior may have appeared similar to many other GRB afterglows. Despite the incredibly high flux and fluence of GRB 080319B, the total jet-corrected observed energy budget ($\sim 4 \times 10^{50}$ ergs) is moderate, and is consistent with the overall distribution for all GRBs[49]. Additionally, if the SSC interpretation of the prompt emission is indeed generic, it implies that a reasonably bright second-order SSC component peaking at ~10-100 GeV may be a common feature in GRBs, and may significantly increase the GRB total energy budget. *GLAST* will soon test this prediction.

15**Table 1 | Observations of GRB 080319B**

| Facility | *Epoch | Band | Peak Flux[††] |
|---|---|---|---|
| *Swift*-BAT | -120 – 182 | 15 – 350 keV | $2.3 \times 10^{-6}$ erg cm$^{-2}$ s$^{-1}$ |
| Konus-Wind | -2 – 230 | 20 – 1160 keV[‡] | $2.3 \times 10^{-5}$ erg cm$^{-2}$ s$^{-1}$ |
| *Swift*-XRT | 67 – $2.5 \times 10^6$ | 0.3 – 10 keV | — |
| Pi of the Sky | -1380 – 468 | White | 5.9 mag |
| TORTORA | -20 – 97 | V | 5.3 mag |
| *Swift*-UVOT | 68 - $10^6$ | white, u, v, b, w1, w2, m2 | — |
| REM | 51 – 2070 | R, I, J, H, Ks | — |
| Liverpool Telescope | $1.8 \times 10^3$ – $2.5 \times 10^3$ | SDSS r,i | — |
| Faulkes Telescope North | $2.5 \times 10^4$ – $2.0 \times 10^5$ | Bessell R,I SDSS r,i | — |
| VLT | 435 – 934 | J, Ks | — |
| Gemini N Photometry | $3.0 \times 10^5$, $4.5 \times 10^5$ | r, i | — |
| HST | $1.6 \times 10^6$ | F606W, F814W | — |
| Gemini N Spectroscopy | $1.2 \times 10^4$ - $1.24 \times 10^4$ | 4100-6800 Å | — |
| HET | $2.0 \times 10^4$ – $2.1 \times 10^4$ | 4100-10500 Å | — |
| Westerbork Synthesis Radio Telescope | $50.5 \times 10^3$ - $2.2 \times 10^6$ | 4.8 GHz | — |
| VLA[†] | $1.98 \times 10^5$ – $2.02 \times 10^5$ | 4.86 GHz | 189 μJy |
| Pairitel[†] | 1.27 – $1.77 \times 10^4$ | J, H, Ks | — |
| KAIT[†] | $1.1 \times 10^3$ – $1.7 \times 10^4$ | Clear, B, V, I | — |
| Nickel[†] | $7.1 \times 10^3$ – $2.4 \times 10^4$ | B, V, R, I | — |
| Gemini S[†] | $8.9 \times 10^4$ – $1.7 \times 10^5$ | g, r, i, z | — |
| Spitzer[†] | $2.20 \times 10^4$ – $2.24 \times 10^4$ | 15.8 μm | — |

Details for our observations and data analysis are given in the SI Methods section.
[†]Observations obtained from external sources as identified in Methods section
*Time since BAT trigger in seconds
[‡]KW light curve measured in 20 – 1160 keV range, peak flux measured in 20 keV – 7 MeV
[††]Peak fluxes listed only if a peak was actually observed

**Table 2 | Summary of Two-Component Jet Parameters**

| | $\alpha_{opt}$ | $\beta_{opt}$ | $\alpha_x$ | $\beta_x$ | p | $\nu_m$ | $\nu_c$ | $t_j$ (s) | $\theta_j^*$ (°) | $E_\gamma$ (ergs) |
|---|---|---|---|---|---|---|---|---|---|---|
| WJRS | 2.49±0.09 | 0.49±0.14 | — | — | — | $>\nu_{opt}$ | $<\nu_{opt}$ $<\nu_x$ | — | — | |
| NJFS | — | — | 1.45±0.05 | 0.76±0.10 | 2.4 | $>\nu_{opt}$ $<\nu_x$ | $>\nu_x$ $<\nu_\gamma$ | $2800^{+900}_{-1400}$ | 4.0 | $2.1 \times 10^{50}$ |
| WJFS | 1.25±0.02 | 0.50±0.07 | 0.95±0.20 | 0.98±0.10 | 2.0 | $<\nu_{opt}$ $>\nu_{opt}$ | $<\nu_x$ | $40.8^{+6.2}_{-9.5} \times 10^3$ | 0.2 | $1.9 \times 10^{50}$ |

WJRS – Wide Jet Reverse Shock
NJFS – Narrow Jet Forward Shock
WJFS – Wide Jet Forward Shock
*Half opening angle

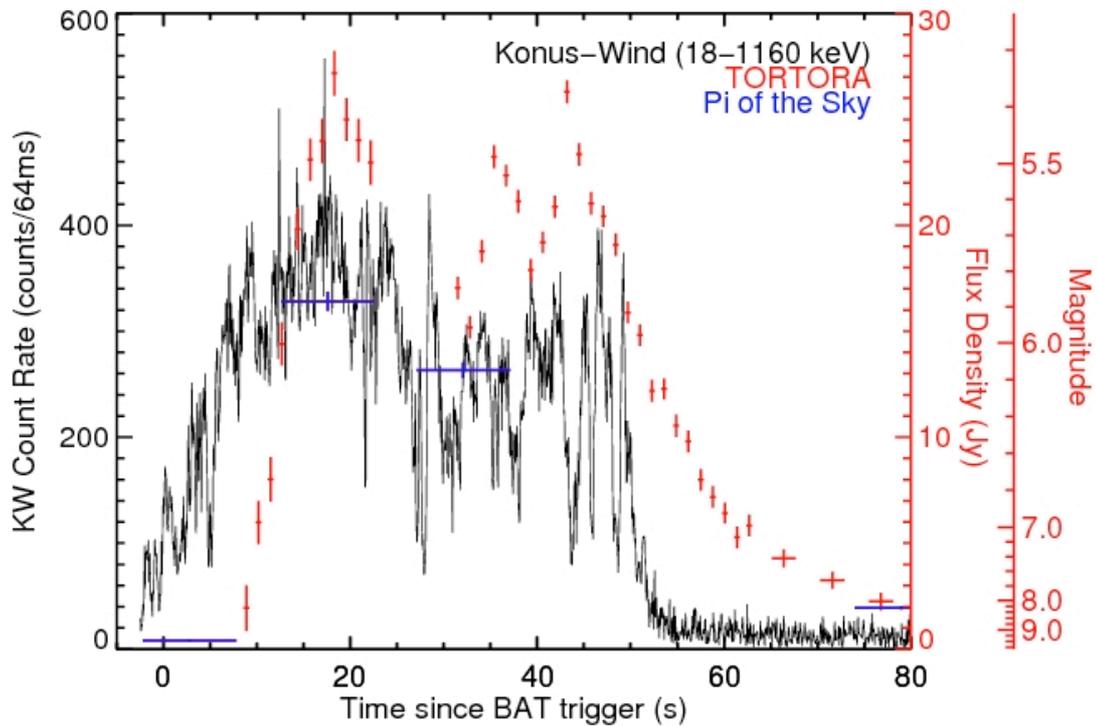

**Figure 1 | Prompt Emission Light Curve**. The Konus-Wind background-subtracted γ-ray lightcurve (black), shown relative to the *Swift* BAT trigger time, $T_0$. Optical data from "Pi of the sky" (blue) and TORTORA (red) are superimposed for comparison. The optical emission begins within seconds of the onset of the burst. The TORTORA data have a gap during the slew of the REM telescope to this field, but show 3 sub-peaks in the optical brightness, reaching a peak brightness of 5.3 magnitudes (white). The γ-ray light curve has multiple short peaks; these are not well correlated with the optical peaks in detail (cf. ref 25), but the optical pulses may be broader and peak somewhat later than the γ-ray pulses, if the optical is slightly below the synchrotron self-absorption frequency, which may account for the lack of detailed correlation. The optical flash, however, begins and ends at approximately the same times as the prompt γ-ray emission, providing strong evidence that both originate at the same site. See Supplementary Materials for more detailed description of correlation tests.

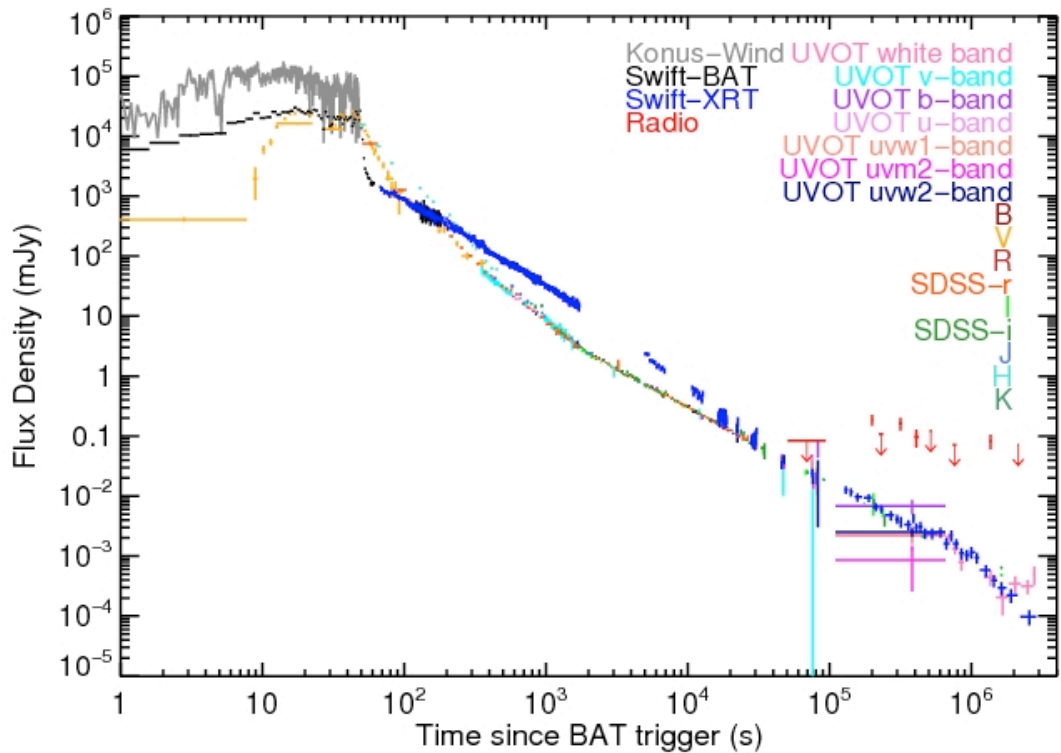

**Figure 2 | Composite Light Curve.** Broadband light curve of GRB 080319B, including radio, NIR, optical, UV, X-ray and γ-ray flux densities. The UV/optical/NIR data are normalized to the UVOT v-band in the interval between $T_0$+500 s and $T_0$+500 ks. The *Swift*-BAT data are extrapolated down into the XRT bandpass (0.3-10 keV) for direct comparison with the XRT data. The combined X-ray and BAT data are scaled up by a factor of 45, and the Konus-Wind data are scaled up by a factor of $10^4$ for comparison with the optical flux densities. This figure includes our own data, plus one VLA radio data point[50], and optical data from KAIT, Nickel, and Gemini-S[22]. The deviations in the NIR points from $T_0$+100 - 600 s are due to strong colour evolution in the SEDs at this time; these points were not included in our overall light curve fits (Supplementary Figure 6).

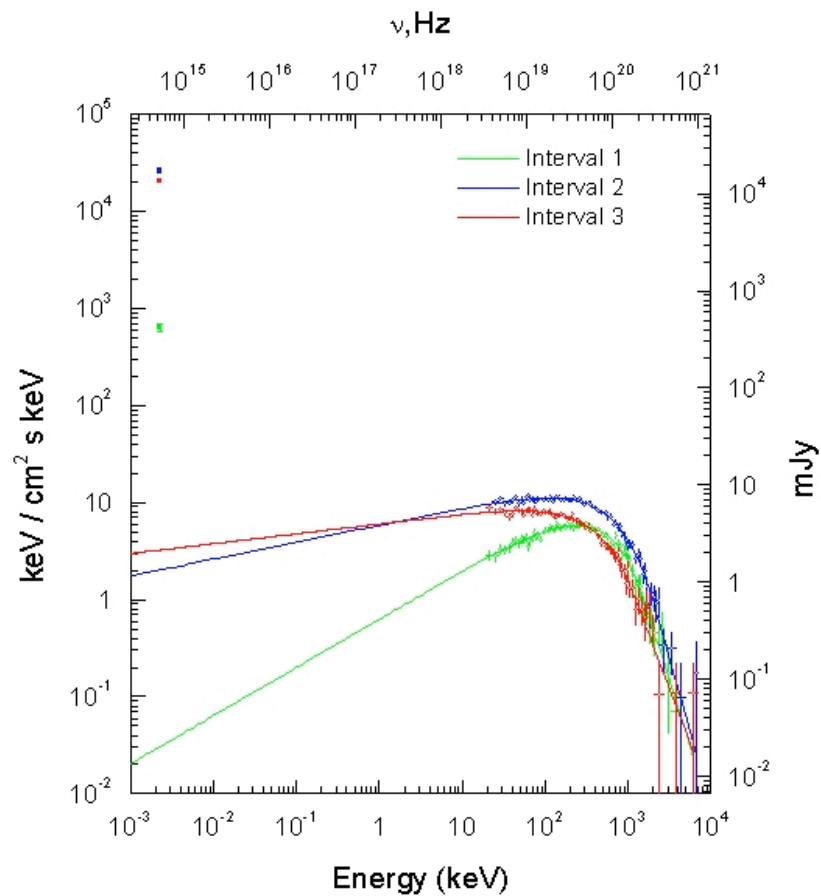

**Figure 3 | Konus-Wind and "Pi of the Sky" Combined Spectra.** Konus-Wind spectra and "Pi of the Sky" flux density in three 10 second time intervals centred at $T_0$+3s, $T_0$+17s, and $T_0$+32s. (Detailed time intervals and γ-ray spectral parameters are given in Supplementary Table 1.) The high energy data points are from Konus-Wind, and the solid line shows the best-fit Band function[16] for each time interval. The low energy points are the "Pi of the sky" flux density measured in approximately the same time interval. The optical flux density exceeds the extrapolation of the γ-ray model by 4 orders of magnitude.

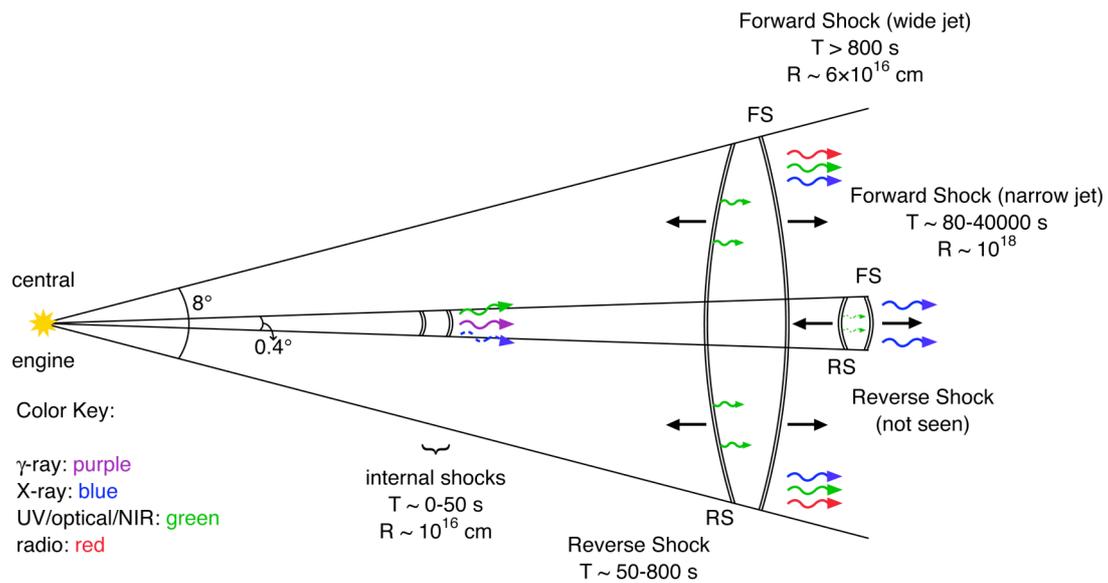

**Figure 4 | Schematic of Two-Component Jet Model.** Summary diagram showing spectral and temporal elements of our two-component jet model. The prompt γ-ray emission is due to the internal shocks in the narrow jet, and the afterglow is a result of the forward and reverse shocks from both the narrow and wide jets. The reverse shock from the narrow jet is too faint to detect compared to the bright wide jet reverse shock and the prompt emission. If X-ray observations had begun earlier, we would have detected X-ray emission during the prompt burst. These expected (but unobserved) emission sources are indicated by the dashed photon lines. Diagram is courtesy of J.D. Myers (NASA).

Supplementary Information accompanies the paper on www.nature.com/nature.

**Acknowledgements** Our research is supported by NASA, National Science Foundation (NSF), Agenzia Spaziale Italiana (ASI), Ministero dell'Universita` e della Ricerca (MUR), Ministero degli Affari Esteri (MAE), Netherlands Organization for Scientific Research (NWO), National Science Foundation of China (NSFC), Science and Technology and Facilities Council (STFC), Slovenian Research Agency, and Ministry for Higher Education, Science, and Technology, Slovenia, and the Polish Ministry of Science and Higher Education.  The Westerbork Synthesis Radio Telescope is operated by ASTRON (Netherlands Insititute for Radio Astronomy) with support from the NWO. The TORTORA team acknowledges Università di Bologna funds "Progetti pluriennali"  that supported the TORTORA camera.  The Liverpool Telescope is operated on the island of La Palma by Liverpool John Moores University in the Spanish Observatorio del Roque de los Muchachos of the Instituto de Astrofisica de Canarias with financial support from the UK STFC. The VLT observatory at Paranal Chile is operated by the European Southern Observatory.  The Faulkes Telescope North is owned and operated by Las Cumbres Observatory Global Telescope Network (LCOGTN), a privately funded, nonprofit organization. The Konus-Wind experiment is supported by the Russian Space Agency and the Russian Foundation for Basic Research. This work is based on observations obtained at the Gemini Observatory, which is operated by the Association of Universities for Research in Astronomy, Inc., under a cooperative agreement with the NSF on behalf of the Gemini partnership: the NSF (United States), the STFC (United Kingdom), the National Research Council (Canada), CONICYT (Chile), the Australian Research Council (Australia), Ministério da Ciência e Tecnologia (Brazil) and SECYT (Argentina), observations made with the NASA/ESA Hubble Space Telescope, obtained at the Space Telescope Science Institute, which is operated by the Association of Universities for Research in Astronomy, Inc. under NASA. We acknowledge the use of public data from the *Swift* data archive.  AJvdH was supported by an appointment to the NASA Postdoctoral Program at NSSTC, administered by Oak Ridge Associated Universities through a contract with NASA.   J.G. gratefully acknowledges a Royal Society Wolfson Research Merit Award.  We thank E. Rol for his helpful comments.

**Author Information** Correspondence and request for materials should be addressed to J.L. Racusin (racusin@astro.psu.edu).

# SUPPLEMENTARY INFORMATION

## METHODS

### γ-ray Observations and Data Reduction

*Swift*-BAT[12] triggered on GRB 080319B 06:12:49 UT on March 19, 2008. The mask-weighted light curve (Supplementary Figure 1) showed a broad flat-topped peak starting at ~ $T_0-10$ sec, ramping up until ~$T_0+10$ sec, then starting to decay at ~ $T_0+50$ sec. It returns nearly to the background level by ~ $T_0+64$ sec at which point a gap in the data from $T_0+64$ to $T_0+120$ and $T_0+302$ to $T_0+660$ occurred due to the onboard data storage buffer filling up, and the event data ended at $T_0+782$. We cannot make detailed spectra during the gaps, but the mask-tagged rate data partially cover the gaps. Given the missing data, $T_{90}$ (15-350 keV) is constrained to be >50 sec. The time-averaged spectrum from $T_0-3.8$ to $T_0+62.2$ and $T_0+120$ to $T_0+151$ sec is fit by a simple power-law model with index of 1.04±0.02. Time resolved spectra shows dramatic softening at ~$T_0+53$ s (i.e. immediately after the bright prompt emission ends; see Supplementary Figure 2). The fluence in the 15-150 keV band is 8.1±0.1×$10^{-5}$ erg cm$^{-2}$. The 1-sec peak flux measured from $T_0+16.87$ sec in the 15-150 keV band is 2.3×$10^{-6}$ erg cm$^{-2}$ sec$^{-1}$. All *Swift* data were analyzed using Version 28 of the Swift software, released with FTOOLs v6.4 (19 November, 2007). The BAT data points are available in the Supplementary Information as an ASCII text file.

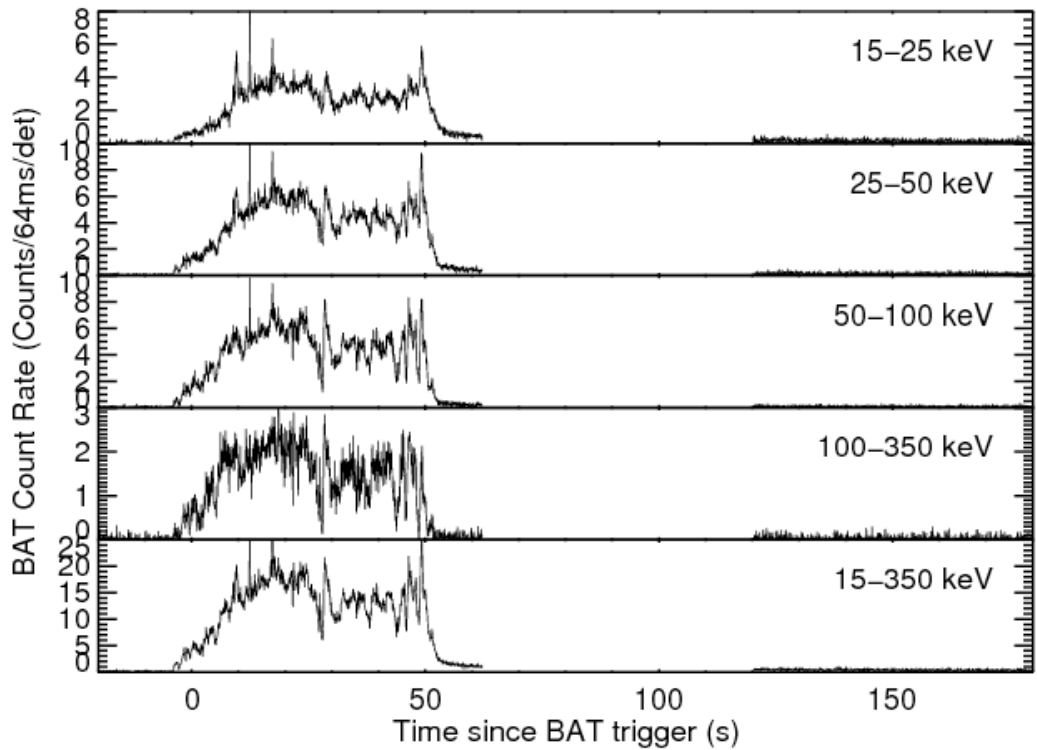

**Supplementary Figure 1 | BAT Mask-weighted Light Curve.** Four channel and combined 64 ms mask-weighted light curve. The spectral softening during the end of the prompt emission is apparent by comparing the intensity of the softer and harder channels.

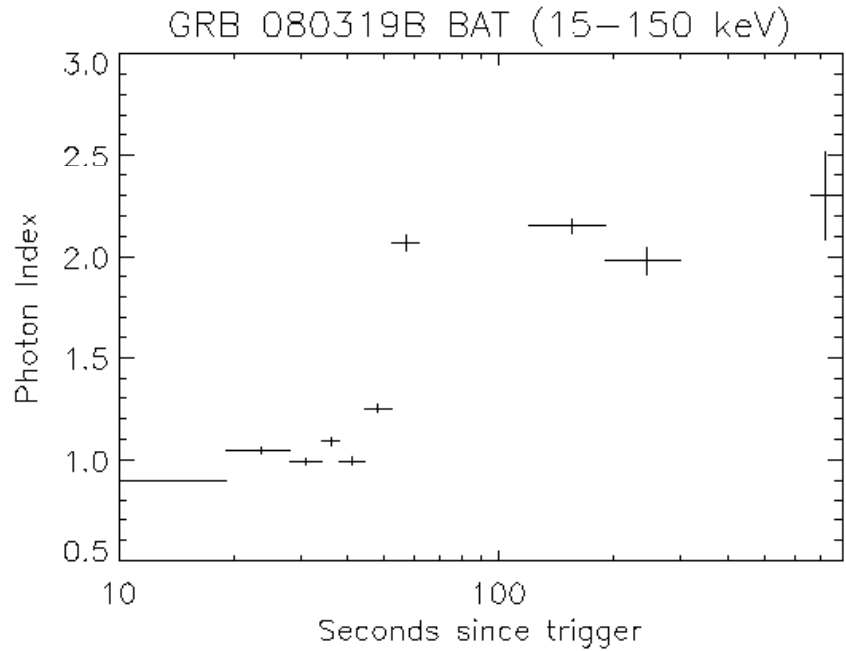

**Supplementary Figure 2 | BAT Spectral Evolution during Prompt Emission.** Time-resolved photon index of single power-law fits to the BAT prompt data. When the main burst ends, the spectrum softens dramatically, with the tail of the burst having a photon index of about 2.1 (slightly steeper than the early XRT photon index).

Konus-Wind (KW)[15] triggered on GRB 080319B at $T_0$(KW)=06:12:50.339 UT, with the S2 detector (which observes the north ecliptic hemisphere) at an incident angle of 42.3 deg. Taking into account the propagation delay, the KW trigger time corresponds to $T_0$(BAT)-1.949 s. KW accumulated 64 spectra in 101 channels (from ~20 keV to ~15 MeV) from $T_0$(KW) to $T_0$(KW) + 174.336 s, on time scales varying from 0.512-2.048 s during the main phase of the burst, to a scale of 8.192 s by the time the signal became undetectable. Data were processed using standard KW analysis tools, and the spectra were fit using XSPEC v11.3. The KW light curves show a complex structure throughout the main pulse ($T<50$ s), followed by a long soft decaying tail detectable up to $T_0$+200 s (Supplementary Figure 3). KW detected strong spectral evolution that is clearly pronounced in the hardness-intensity correlation (Supplementary Figure 3). The KW data points are available in the Supplementary Information as an ASCII text file.

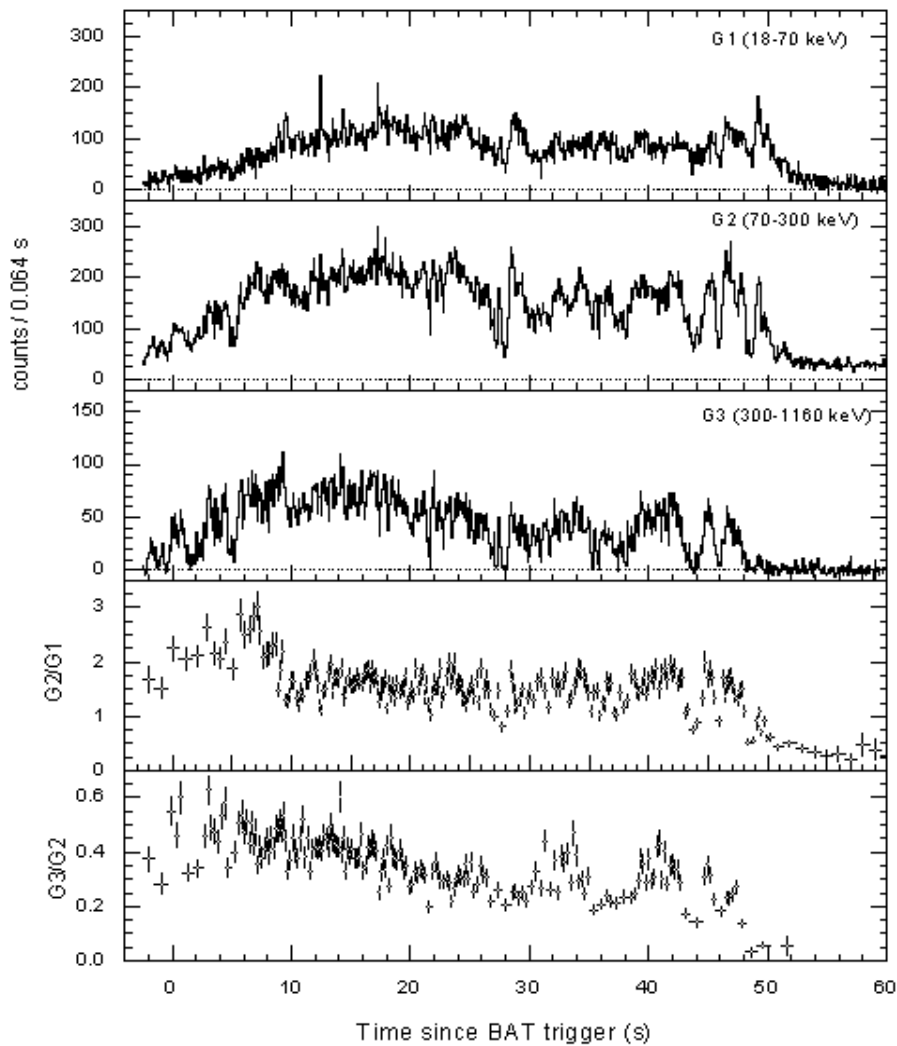

**Supplementary Figure 3 | Konus-Wind Multi-Band Light Curve, and Hardness Ratios.** Multi-band KW light curves (panels 1-3) show early hard to late soft structure transitions consistent with the hardness ratios (bottom 2 panels).

**X-ray Observations and Data Reduction**

*Swift*-XRT[24] began observing GRB 080319B at $T_0+51$ s. The Level 0 event data were downloaded from the Swift Data Centre. The source was extremely bright, and heavily piled-up even in Windowed Timing (WT) mode. The pileup effect[51] is a result of multiple photons striking within 3 adjacent pixels within the 1.8 ms between the detector readouts. To correct for the pileup effects, we estimated the region around the centre of the source that needs to be excluded for a particular count rate, by extracting and fitting spectra to absorbed power-laws with different exclusion regions, until excluding additional pixels did not change the spectral index (using the method described by ref. 52). After the piled-up events were removed, we had to estimate the fraction of the flux removed by excluding the centre of the PSF. We accomplished this by fitting spectra with ancillary response files (ARFs), made with the task *xrtmkarf*, with and without PSF and exposure map corrections. By comparing the ratio of the unabsorbed fluxes between the spectra fit with the corrected and uncorrected ARFs, we determined the pileup correction factor to apply to the individual count rate intervals

4within the light curve. In the brightest uncorrected count rate regime (1200-1400 cts/s), we removed the central 16 pixels from the source for light curve and spectra construction. The exclusion region was reduced as appropriate for lower count rates. The light curve was constructed manually for the WT data and the Photon Counting (PC) data before $T_0+10^5$s using the tools *xrtproducts* and *xrtlccorr*. The late-time XRT light curve was extracted from the XRT team light curve repository tools[53] and rebinned for higher signal-to-noise. The XRT data points are available in the Supplementary Information as an ASCII text file.

**Optical/UV/IR Photometric Observations and Data Reduction**

*Swift*-UVOT[23] began observing GRB 080319B $T_0+68$ s. The source was so bright that UVOT suffered from heavy coincidence loss and scattered light when the observations began. The v-band data before $T_0+350$s, and the white data before $T+1000$s could not be recovered. The later data were processed using the standard *Swift* software tool *uvotmaghist*. We extracted counts using a circular aperture with a radius of 5" when the count rate was above 0.5 counts s$^{-1}$, and 3" aperture when the count rate had dropped below 0.5 counts s$^{-1}$, and an appropriate background region. We applied coincidence loss corrections and standard photometric calibrations[54]. The UVOT data points are available in the Supplementary Information as an ASCII text file.

"Pi of the Sky"[18] consists of two CCD cameras with 2000×2000 pixels and photo lenses with 71 mm diameter and 85 mm focal length. The field of view (FoV) is 20°×20° with the scale of 36 arcsec/pixel. An IR-cut filter passing 390-690 nm is used to reduce the sky background setting the limiting magnitude at 11.5-13.0 for single exposures and 12.5-14.0 for 10 co-added images, depending mainly on Moon phase and position. The FoV is continuously monitored with 10s exposures as the data are analyzed by algorithms searching for fast optical transients. The position of GRB 080319B was observed by "Pi of the Sky" from 5:49 UT, i.e. 23 minutes before the burst[17]. The first image with the flash visible started 2.3 s before the Swift trigger GRB 080319B. The flash was registered near the border of the frame. After taking 3 images, the alert from *Swift* was received and the apparatus moved to position the GRB in the centre of the frame accounting for a gap in the data between 37 s and 73 s. After peaking at 5.87 mag subsequent images show the decay of the source below 12$^{th}$ mag after 5 minutes. The "Pi of the Sky" data points are available in the Supplementary Information as an ASCII text file.

TORTORA[19] is an autonomous wide-field camera with a 30°×24° FoV mounted on the side of the REM telescope. The camera consists of a 120mm objective (1:1.2), an image intensifier reducing the linear field size by a factor of 4.5, and a fast TV-CCD, which allows it to take images of a large area of the sky (about 600 sq.deg.) with a frame rate of 7.5 frames per second (exposure time of 0.13 s) without gaps between exposures. The camera software is able to perform data processing, detection and classification of optical transients in real time[55]. TORTORA was already observing the field of GRB 080319B 20s before the BAT trigger and detected emission from the GRB starting from about 8s after the burst, when the optical counterpart become brighter than V~8. The observations were carried out at high zenith angle with part of the field of view obscured by the REM dome. TORTORA was observing in white light and sensitivity is driven by the S25 photocathode. Data have been calibrated to the Johnsson V system using several nearby Tycho2 catalogue stars[56]. From ~ $T_0+23-30$ s,

TORTORA was not able to collect data as REM was slewing to the field of GRB 080319B after having received the *Swift*-BAT alert[14]. The TORTORA data points are available in the Supplementary Information as an ASCII text file.

REM[20] began observing the afterglow of GRB 080319B in the H and then R filters starting ~50 s after the burst until the field was too low on the local horizon[58]. Later observations were also carried out in I, J, and Ks filters. Optical data were calibrated by observing Landolt standard stars, while NIR data were calibrated with respect to the 2MASS catalogue[57]. The REM data points are available in the Supplementary Information as an ASCII text file.

The Liverpool Telescope (LT) and Faulkes Telescope North (FTN) run a common robotic GRB response system[59], which automatically responds to GRB alerts on the GCN socket. The LT however was already observing 080319A and did not respond to GRB 080319B until 30 minutes after the burst time. The geographic separation of the two telescopes allows them to coordinate to provide nearly 24-hour coverage and r, i-band monitoring was continued until three days after the burst time, when it was too faint for the 2m aperture telescopes. The imaging data were reduced through the standard automated CCD reduction pipeline. Most of the observations used SDSS-like r, i filters and magnitude zero points were derived with respect to four nearby comparison stars drawn from the SDSS DR6 online catalogue. For the few Bessell R and I observations we used the R2 and I magnitude from the USNO-B1 catalogue and same four comparison stars. The LT and FTN data points are available in the Supplementary Information as an ASCII text file.

VLT obtained early observations while equipped with the ISAAC NIR camera in J and Ks bands calibrated with respect to the 2MASS catalogue. Magnitudes were computed by means of aperture photometry or PSF fitting techniques when required. The VLT data points are available in the Supplementary Information as an ASCII text file.

Gemini N observations were obtained in r and i band using the GMOS instrument on March 22 14:44 UT, with an exposure of 5×200s in each filter. Another epoch (5x100s) in both filters was obtained on March 24 11:29 UT. Photometry was calibrated to AB magnitudes via SDSS stars in the field. The Gemini-N data points are available in the Supplementary Information as an ASCII text file.

Pairitel, KAIT, Nickel, and Gemini S observations were obtained from (ref. 22), however Pairitel data were not used in any of the light curves due to discrepancies with other simultaneous data from our group (possibly due to early saturation) and unreasonably small magnitude errors (~milli-magnitude)[22]. Some Pairitel data were however used in the SEDs with a systematic error term added.

Spitzer points were obtained from GCN circulars[60].

HST observed the field of GRB 080319B on April 7th, using WFPC2 to obtain 8x400s exposures in both F606W and F814W filters. Calibration used standard HST zero-points and CTE corrections, and was transformed to r and i AB-magnitudes assuming a power-law SED. The HST data points are available in the Supplementary Information as an ASCII text file.

**Optical/NIR Spectroscopic Observations and Data Reduction**
Gemini N Spectroscopy was obtained with GMOS beginning 9.28 am UT. Four 10 minute exposures were obtained with the B600 grism and 0.75 arcsec slit. This spectrum (shown in Supplementary Figure 4) shows two absorption systems; the highest redshift, z=0.937, presumably corresponds to the host galaxy.

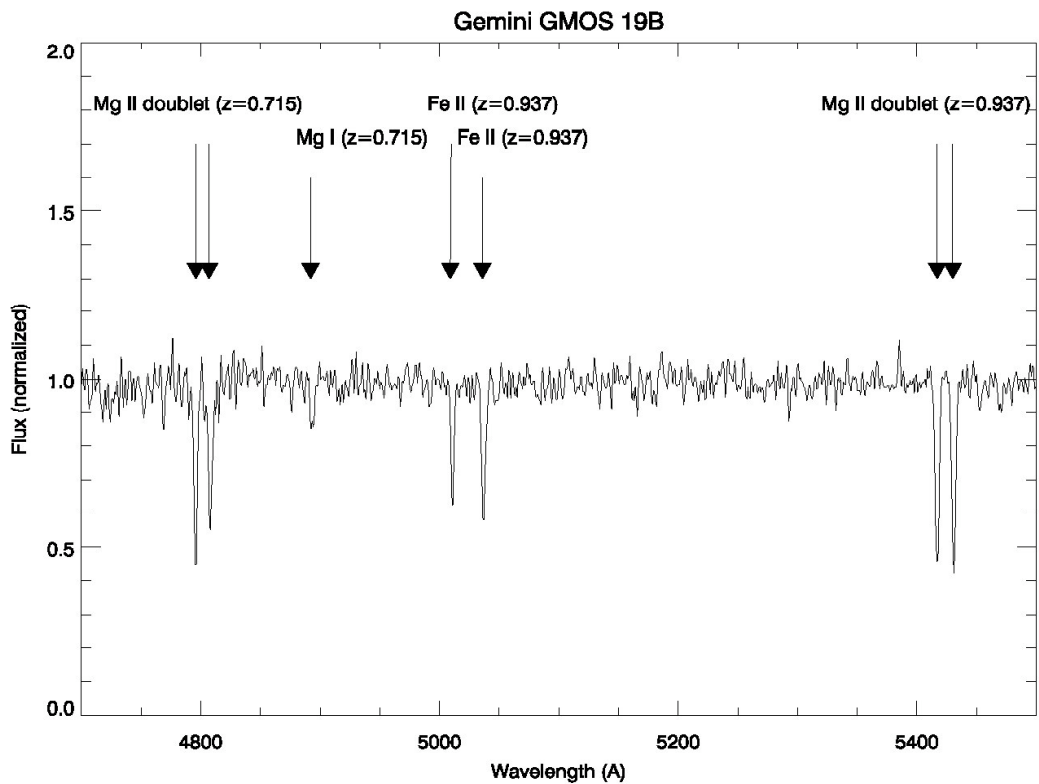

**Supplementary Figure 4 | Gemini-N Spectrum.** Small section of the Gemini-N spectrum taken ~3 hours after the BAT trigger. Two prominent absorption systems are present at z=0.715 and z=0.937.

HET Spectroscopy was taken with the Marcario LRS spectrograph (R ~230), ~6 hours after the BAT trigger, with a 1200 second integration time. The reduction was done using the standard IRAF package tools for image de-biasing, flat-fielding, and wavelength calibration. Cosmic-ray rejection has been performed with the *Lacos_spec* routine[61]. The 1d spectrum (shown in Supplementary Figure 5) was extracted using the IRAF *apall* task, and was flux calibrated using a HZ44 spectro-photometric standard observation taken the same night. We clearly detected MgII doublets at the GRB redshift and some other metal absorption features due to intervening system along the line of sight.

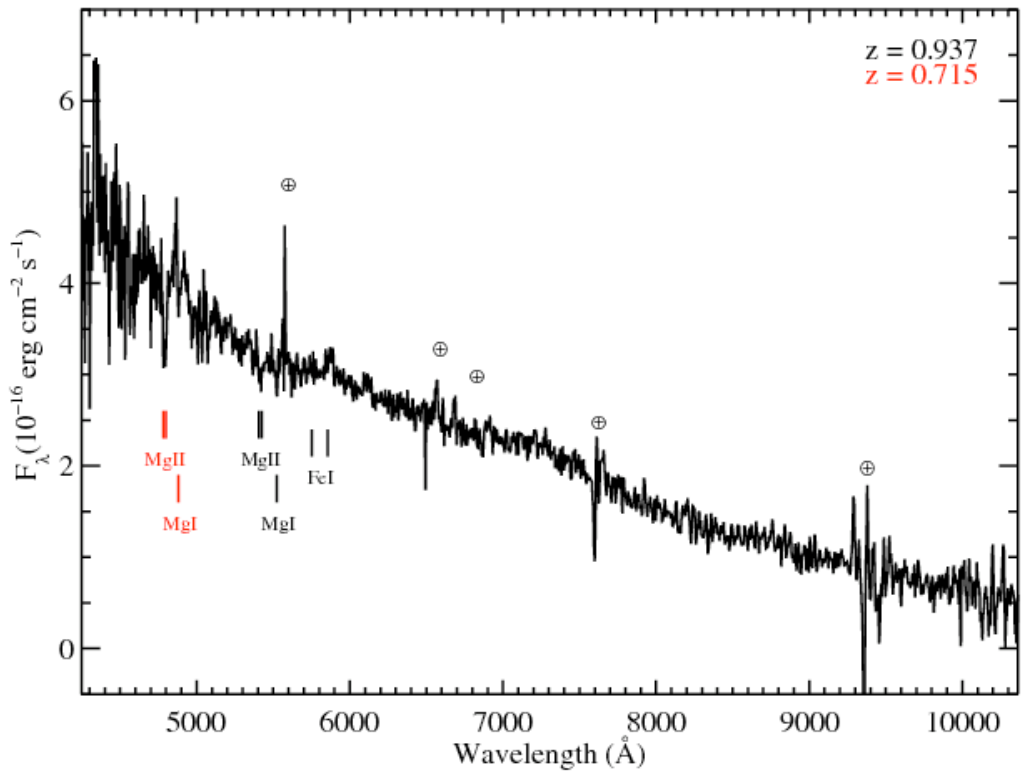

**Supplementary Figure 5 | HET Spectrum.** Complete HET spectrum taken ~6 hours after the BAT trigger showing 2 prominent absorption systems at z=0.715, 0.937. The ⊕ symbol indicates atmospheric lines.

**Radio Observations and Data Reduction**

Radio observations were performed with the Westerbork Synthesis Radio Telescope (WSRT) at 4.9 GHz, using the Multi Frequency Front Ends[62] in combination with the IVC+DZB back end in continuum mode, with a bandwidth of 8x20 MHz. Gain and phase calibrations were performed with the calibrator 3C286. Reduction and analysis were performed using the MIRIAD software package. Radio observations with WSRT from 0.6-1.1 days resulted in an upper limit of $F_{4.86\ GHz} < 84$ μJy. VLA observations showed a peak of $F_{4.86\ GHz} = 189\pm39$ μJy at 2.31 days[49]. Subsequent observations by WSRT over the following weeks showed that the radio afterglow has large variations in flux density between epochs. We attribute this to the effect of Galactic scintillation[47,48] which is especially strong while the source is physically very small in the earliest observations; therefore, we cannot measure the shape of the radio afterglow decay or use the early data in our SEDs, however the existence of the emission and the flux at late times (after the source size is large enough for scintillation to subside) are consistent with the forward shock of the wide jet propagating into a stellar wind density distribution. The WSRT data points are available in the Supplementary Information as an ASCII text file.

VLA[49] data were obtained from GCN circulars.

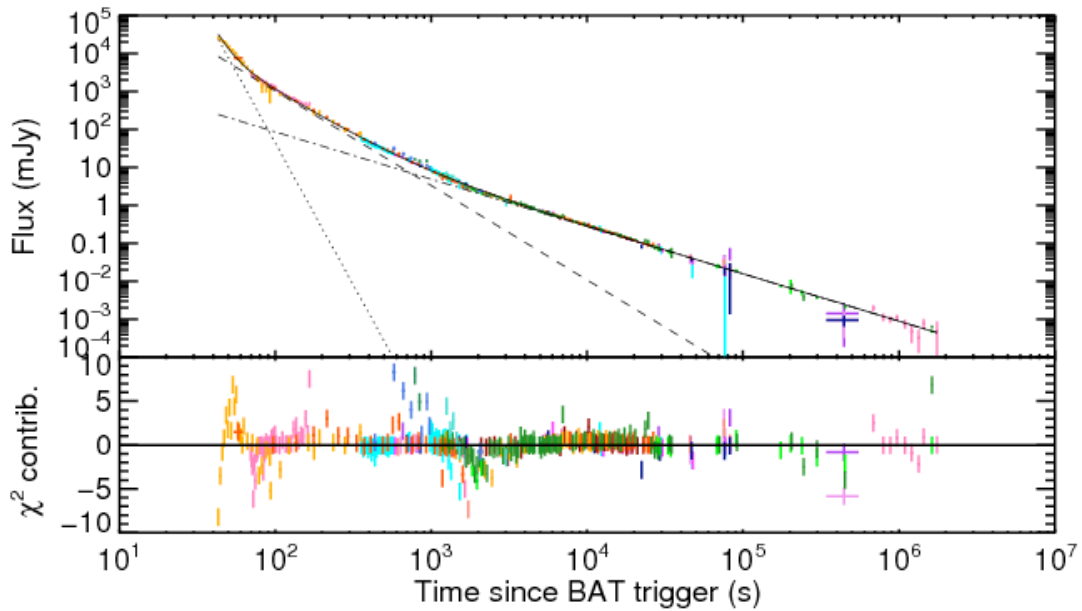

**Supplementary Figure 6 | Three-Spectral Component Fit to the Decaying Optical Transient** Following the peak of the prompt optical flash, the optical transient light curve displays three distinct components that dominate in the intervals t<50s, 50s<t<800s, and t>800s. The initial decay of the bright optical flash is a power-law with $\alpha_1$=6.5±0.9 (dotted line). This is superimposed on a power-law with decay index $\alpha_2$=2.49±0.09 (dashed line) that dominates in the middle time interval and a third power-law with $\alpha_3$=1.25±0.02 (dot-dashed line) that dominates at late times. The residuals show some evidence for bumps superposed on top of this general decay (see also ref. 22). We have added a 7% relative systematic error to all data points in order to account for calibration uncertainties. The data are colour-coded in the same scheme as Figure 2.

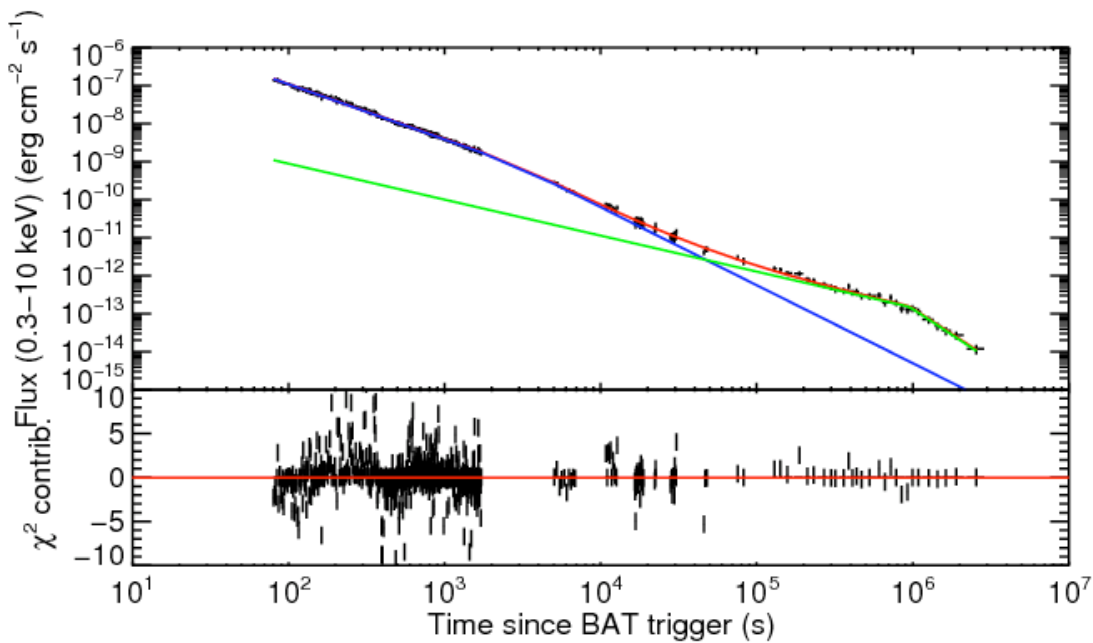

**Supplementary Figure 7 | Two-Component Jet Model fit to X-ray Afterglow.**
The X-ray afterglow is best described by the superposition of two broken power-laws, which is consistent with the narrow and wide jets of a two-component jet expanding into a stratified wind environment. The narrow jet dominates the first ~40 ks of the afterglow as indicated by the blue line, which shows the fit to the narrow jet component. After the narrow jet break decays, the wide jet dominates as indicated by the green line fit to late afterglow. The red line shows the superposition of both components and the overall fit to the X-ray light curve.

# BROADBAND MODELLING

Combining multi-band data to probe the evolution of the spectral components is key to understanding the wider picture of GRB 080319B. We used several different methods to understand the relationship between the optical, X-ray and γ-ray data in different temporal regimes. We show that the prompt optical and γ-ray emission is from the same physical region but different spectral components by comparing and testing the correlation between them.

In Figure 2, we show that the optical emission is several orders of magnitude above the extrapolation of the γ-ray emission during the prompt phase. KW data taken from several time intervals that overlap with "Pi of the Sky" intervals, were fit to Band functions, and those parameters are given in Supplementary Table 1. The optical and γ-ray data cannot be accounted for together within the Band function framework, therefore the optical emission must stem from an additional spectral component.

**Supplementary Table 1 | Konus-Wind Band Function Spectral Fits**

| $t_{start}$ (s) | $t_{stop}$ (s) | a | b | $E_p$ (keV) | $\chi^2$/dof |
|---|---|---|---|---|---|
| -2 | 8 | $-0.504^{+0.039}_{-0.038}$ | $-3.208$* | $729^{+34}_{-32}$ | 97.8/82 |
| 12 | 22 | $-0.826^{+0.022}_{-0.021}$ | $-3.426$* | $751\pm26$ | 83.2/81 |
| 26 | 36 | $-0.898^{+0.031}_{-0.029}$ | $-3.270^{+0.384}_{-1.061}$ | $537^{+28}_{-27}$ | 74.0/82 |

*90% confidence upper limit

At a first glance, the γ-ray and optical data appear to be generally correlated. To quantify this correlation, we compared the time-resolved BAT data in 4 energy bands to the TORTORA data in those same time intervals (Supplementary Figure 8). The strengths of the correlations (Supplementary Figure 9) were quantified using a linear Pearson correlation function, and the Spearman and Kendall rank-correlation functions. All three measurements indicate a correlation during the initial rising and final declining portions of the prompt emissions, and no correlation in the central portion. The rising and declining trends are not surprising given that the correlation measures the general similar duration of the light curves. The fact that the central brightest portion of the emission is not correlated shows that the lack of correlation between the V band and γ-ray data is genuine.

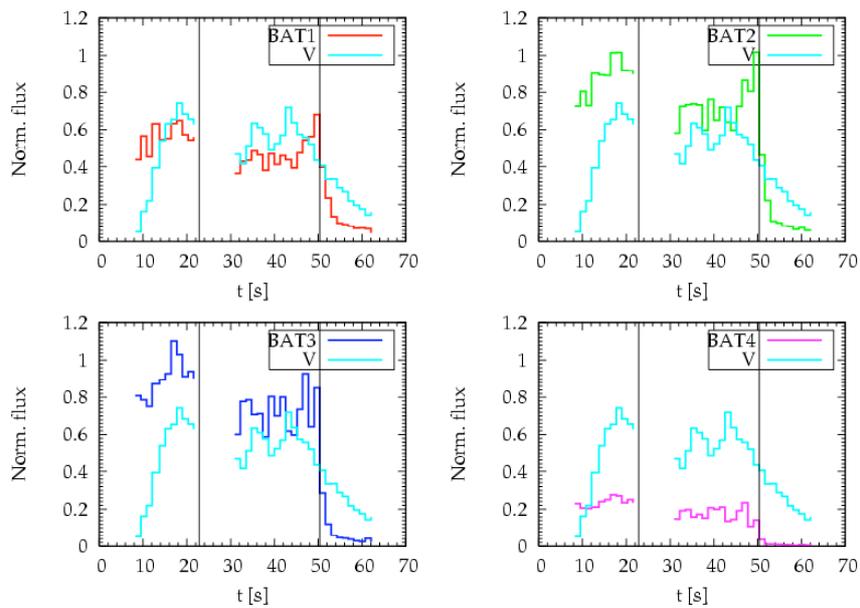

**Supplementary Figure 8 | γ-ray – Optical Prompt Emission Correlation Test Time Intervals.** The division of the prompt γ-ray emission in the 4 BAT energy bands (15-25, 25-50, 50-100, 100-350 keV) and the V-band TORTORA optical emission split into three time intervals indicating the rising, middle, and decaying portions.

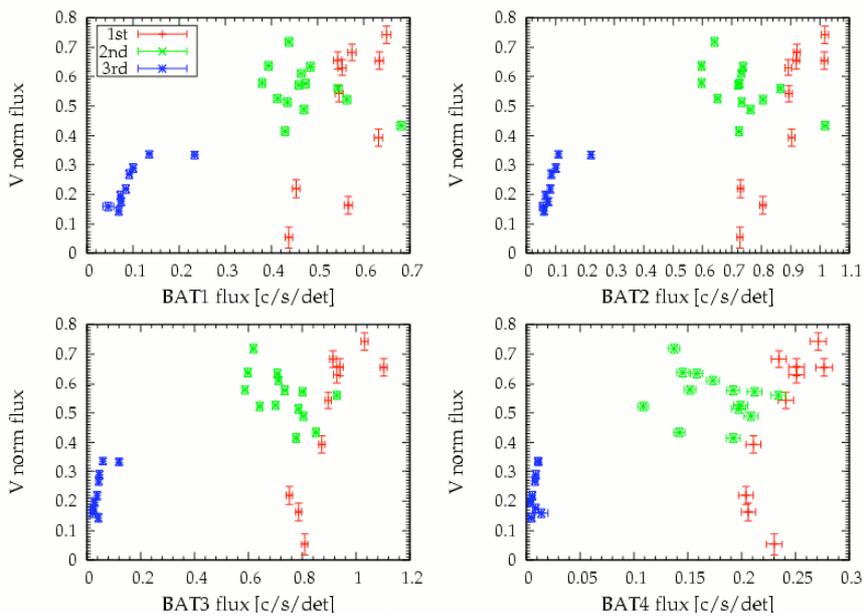

**Supplementary Figure 9 | γ-ray – Optical Prompt Emission Correlations**
The correlation between optical and γ-ray data points in the time intervals and 4 BAT energy bands indicated in Supplementary Figure 8. There is a strong correlation in the rising (blue) and decaying (red) portions of the light curves in all four bands, but no measurable correlation in the middle portion (green), as shown in the bottom 4 panels. This is in contrast to the conclusions of ref. 25.

**Two-Component Jet Break SED Modelling**

Our most plausible model for explaining the spectral and temporal evolution of GRB 080319B, involves a two-component jet with a very narrow (0.2°) highly relativistic core, surrounded by a wide (4°) jet with more typical parameters expanding into a wind-like environment. The two-component jet model was derived from examination of the unusually shaped X-ray light curve and fits to the closure relations. We explain the X-ray light curve as the superposition of two broken-power laws (Supplementary Figure 7), indicating the narrow jet forward shock with its post-jet break decay, and the wide jet with its post-jet break decay. The optical light curve is describes by the super-position of 3 power-laws, with the first indicating the tail of the prompt emission, the second indicating the reverse shock from the wide jet, and the third power-law is the forward shock of the wide jet. The optical emission from the narrow jet is never observed because it is overwhelmed by the wide jet bright reverse shock emission. This model implies the existence of 4 distinct spectral components, 3 of which are visible: the narrow jet reverse shock (NJFS), the narrow jet forward shock (NJFS), the wide jet reverse shock (WJRS), and the wide jet forward shock (WJFS). Each of these spectral components follows the double broken power-law form of the synchrotron spectrum[6,63], each with a moving $\nu_m$ and $\nu_c$, a particular electron spectral index ($p$), and fading at a distinct rate. They dominate at different times and frequencies throughout the SED evolution.

The closure relations[5,63] adequately describe the very early and late temporal and spectral behaviour specific to our model. However, to show that our model works during the intermediate times, we characterize the evolution of the spectral components when they are super-imposed and cannot be cleanly separated. To do this, we constructed SEDs at $T_0$+150 s, 250 s, 350 s, 720 s, 1500 s, 5856 s, $10^4$ s, $3\times10^4$ s, $8\times10^4$ s, $2\times10^5$ s, and $5\times10^5$ s. Rather than assume that we can disentangle these components, we independently fit the extinction at each SED epoch, finding a mean value of E(B-V)=0.05. We apply this fixed mean extinction and fit the X-ray absorption, removing these effects from the SEDs before they are input into our modelling. At early times, the WJRS dominates the optical emission, and the NJFS dominates the X-rays. Given these assumptions, we can explain the flattening of the SEDs in the mid-times, as the various spectral frequencies of the different components move with time following the synchrotron scaling laws (Supplementary Figure 10). We do not attempt joint numerical fitting of the SEDs in this work, but rather fit or approximate various parameters when they are cleanly identifiable at times when only one component dominates one spectral band, and then scale them appropriately. The normalizations of the spectral components in the V-band and "X-band"(=2 keV) are scaled with the light curve temporal decays.

This model is not a fit to the data set, but rather with the appropriate parameter choices, we can reproduce the SEDs reasonably well. When looking in detail at each SED, one can clearly see that the models are not perfect fits to the data, but demonstrate that with enough leeway it is possible to qualitatively reproduce the shapes. We note that we get a much better fit if we slow down the temporal dependence of $\nu_m$ to $\sim t^{-1.1}$ rather than the normal $t^{-1.5}$ that is expected for the Wind models. This is necessary because $\nu_m$ of the narrow jet starts at the soft end of the X-ray band and cannot pass through the optical until it gets faint enough, or it produces a bump in the optical band that is not observed. The SED models with $v_m \sim t^{-1.1}$ are shown in Supplementary Figure 12. Given a two-component jet, all of these model components are expected, and they must add together similarly to the way we have described. We note that a synchrotron spectrum with smooth breaks[63] might provide better fits than the simple broken power-laws that we have used here, but this is beyond the scope of this work.

In order to ascertain the conditions describing the optical and X-ray afterglows, we apply the closure relations[5,63] which describe the temporal and spectral evolution of GRBs, depending on environment, cooling regime, spectral regime, and jet properties. The closure relation for the forward shock of the narrow jet predicts $(\alpha_x - 1.5\beta_x) = 0.5$, marginally consistent with the observed value $(\alpha_{x,1} - 1.5\beta_{x,1}) = 0.31 \pm 0.16$ and $p \sim 2.4$, where the parameter $p$ describes the index of the assumed power-law distribution of electrons accelerated in the relativistic shock. The electron spectral index derived from the late optical decay index is $p = (4\alpha_{opt,3}+1)/3 \sim 2.0$. Once the narrow jet component faded below the wide jet component, the emission is best described if the cooling frequency is below the X-ray band ($v_c, v_m < v_x$), for which case we expect $(\alpha_x - 1.5\beta_x) = -0.5$, consistent with the observed $\alpha_{x,3} - 1.5\beta_{x,3} = -0.52 \pm 0.25$. The electron spectral index for the wide jet estimated from the late-time X-ray decay index is $p = (4\alpha_{x,3}+2)/3 = 1.93 \pm 0.27$, consistent with the optical results and with the X-ray spectral index at late times.

A wind environment for GRB 080319B is supported by the following facts: (1) When applying the closure relations to the early and late X-ray afterglow (60 s $< t <$ 2800 s, $4 \times 10^4$ s $< t$), and the late optical afterglow ($t >$ 1000 s), the wind model is preferred; (2) in the forward shock wind model, the flux density at $v_m$ is $F_{v_m} \propto v_m^{1/3}$. Since $v_m \propto t^{-3/2}$, the crossing time of the observing frequency $v$ by $v_m$ is $t_m \propto v^{-2/3}$, where $v$ can be 4.8 GHz (radio), $4.7 \times 10^{14}$ Hz (optical), and $4.8 \times 10^{17}$ Hz (X-ray). The optical $t_m \lesssim 10^3$ s, therefore we expect $t_m \lesssim 10$ s for the X-ray afterglow and $t_m \lesssim 2 \times 10^6$ s for the radio afterglow. If we take $F_{v_m} \sim 0.2$ mJy in the radio band, then we obtain $F_{v_m} \sim 100$ mJy in the X-ray band and $\sim 10$ mJy in the optical band, which are consistent with observations.

The late X-ray ($t > 4 \times 10^4$ s), optical ($t \gtrsim 10^3$ s), and radio afterglow thus can be explained as the forward shock emission of the wide jet, with the electron energy index $p \sim 2.0$ inferred from the temporal and spectral indices. The X-ray afterglow decays faster than the optical with a larger spectral index impling $v_{opt} < v_{cm} < v_x$, where $v_{cm} = 4.2 \times 10^{12} \varepsilon_{e,-0.5}^{-1/4} \varepsilon_{B,-2.5}^{-1/4} \zeta_p^{1/2} E_{53}^{1/2} A_*^{-3/2}$ Hz is the critical frequency when $v_c = v_m$ (ref. 64), and $\zeta_p = 6(p-2)/(p-1) < 1$. A tenuous wind ($A_* \ll 1$) is thus required. To fit the X-ray (2 keV) flux density of 0.1 μJy at $t=3 \times 10^6$ s ($v_m < v_c < v_x$), the optical flux density of $\sim 3$ μJy at $t=2.55$ days ($v_m < v_{opt} < v_c$), and the radio flux density of $\sim 150$ μJy at $t \sim 3$ days (4.8 GHz $< v_m < v_c$), we have $A_* \sim 2.1 \times 10^{-3} E_{53} \varepsilon_{e,-0.5}^{-1}$, $\zeta_p \sim 0.04 E_{53}^{1/2} \varepsilon_{e,-0.5}^{-3/2}$, and $\varepsilon_B \sim 8.0 \times 10^{-3} E_{53}^{-3} \varepsilon_{e,-0.5}^2$. Combining $\zeta_p < 1$ and an empirical $\varepsilon_B \gtrsim 10^{-5}$, an upper limit for $A_* \sim 3 \times 10^{-2}$, is obtained. Although not tightly constrained, a set of reasonable parameter values for the wide jet can be $E_{k,iso} \sim 10^{53}$ erg, $A_* \sim 10^{-2}$, $\varepsilon_e \sim 0.07$, $\varepsilon_B \sim 3 \times 10^{-3}$, and $p \sim 2.07$. If the X-ray light curve break at $t \sim 11$ days is a jet break, then we can estimate the half opening angle of the wide jet as $\theta_j \sim 0.064 \sim 4°$ and the true jet energy $E_j \simeq E_{k,iso}\theta_j^2 \sim 1.9\times 10^{50}$ erg. The source angle of the late afterglow is $\theta_s = (1+z)^2 R/(\Gamma D_L) = 3.1(t/\text{day})^{3/4}$ micro-arcsec. Given the scattering measure $SM = 1.053\times 10^{-4}$ kpc m$^{-20/3}$ in the direction of this GRB, the condition for strong scattering is $\nu < 10.4\, SM_{-3.5}^{6/17} d_{\text{scr,kpc}}^{5/17} = 7.06$ GHz, and the early flux density at 4.8 GHz is modulated by scintillation.

The early X-ray afterglow ($t<4\times 10^4$ s) can be explained within the context of a narrow jet, where the break at 3 ks is the consequence of a very early jet break. The narrow jet and the wide hollow jet are coaxial and observed within the cone of the narrow jet. The prompt γ-ray burst and optical flash came from internal shocks within the narrow jet. After the internal shock phase, the narrow jet propagated into the stellar wind and drove an external shock. According to the closure relations, we have $\nu_m<\nu_X<\nu_c$ from the beginning of XRT detection, and the electron energy index $p\sim 2.4$ for the wide jet. The possibility of $\nu_X>\nu_{cm}$ is too small (in this case one would obtain $\varepsilon_e < 4\times 10^{-5}$ when modelling the observations), therefore hereafter we focus on the case of $\nu_{opt}<\nu_X<\nu_{cm}$. The following several additional constraints on the parameters of the narrow jet are: (1) the optical flux density at $t_m$ cannot exceed the observed value $F_\nu \sim 7(t/2.55\text{ days})^{-1.23}\mu$Jy, which means the optical flux density cannot exceed the optical flux density from the wide at $t_m$ ;(2) X-ray (2 keV) flux density ~0.8 mJy at $t=10^3$ s; (3) the time when $\nu_m$ crosses the X-ray band must be earlier than $t=60$ s. Combining the above constraints, we get $\varepsilon_{e,-0.5}^{1.4}\varepsilon_{B,2.5}^{0.85} E_{53}^{0.85} A_* \sim 5.4\times 10^{-4}$, $1.1\times 10^{-3} A_{*,-2}^{2/5} \le \varepsilon_e \le 0.23 A_{*,-2}^{-1/10.6}$ and $\varepsilon_{e,-0.5}^{1.85}\varepsilon_{B,-2.5}^{1.03} E_{53}^{1.78} \ge 2.3$. A set of reasonable parameter values of the narrow jet can be $E_{k,iso}\sim 3.5\times 10^{55}$ erg, $A_* \sim 10^{-2}, \varepsilon_e \sim 0.2, \varepsilon_B \sim 5.7\times 10^{-7}$, and $p\sim 2.4$. Taking the observed jet break time $t_j=2800$ s for this narrow jet, we can get the half-opening angle of the narrow jet as $\theta_j = 3.4\times 10^{-3} \sim 0.2°$ and the total jet energy of $E_j=2.1\times 10^{50}$ erg.

With the above preliminary results on the model parameters, we can differentiate from which jet the reverse shock is responsible for the optical emission during 80 s $<t<1000$ s. Using the scaling law for the forward shock radius, we can estimate the radius when the reverse shock crosses the shell ($t=t_\times \sim 60$s), i.e., $R_\times = 5.7\times 10^{15} E_{53}^{1/2} A_*^{-1/2}$ cm. The duration of the high latitude emission is $(1+z)R_\times\theta_j^2/2c$, which is ~7070 s for the wide jet and ~445 s for the narrow jet. Therefore, the reverse shock of the wide jet produced the optical emission in the intermediate times. Further details of this model will be presented in a subsequent paper (Wu, X. F. et al. (2008), in preparation).

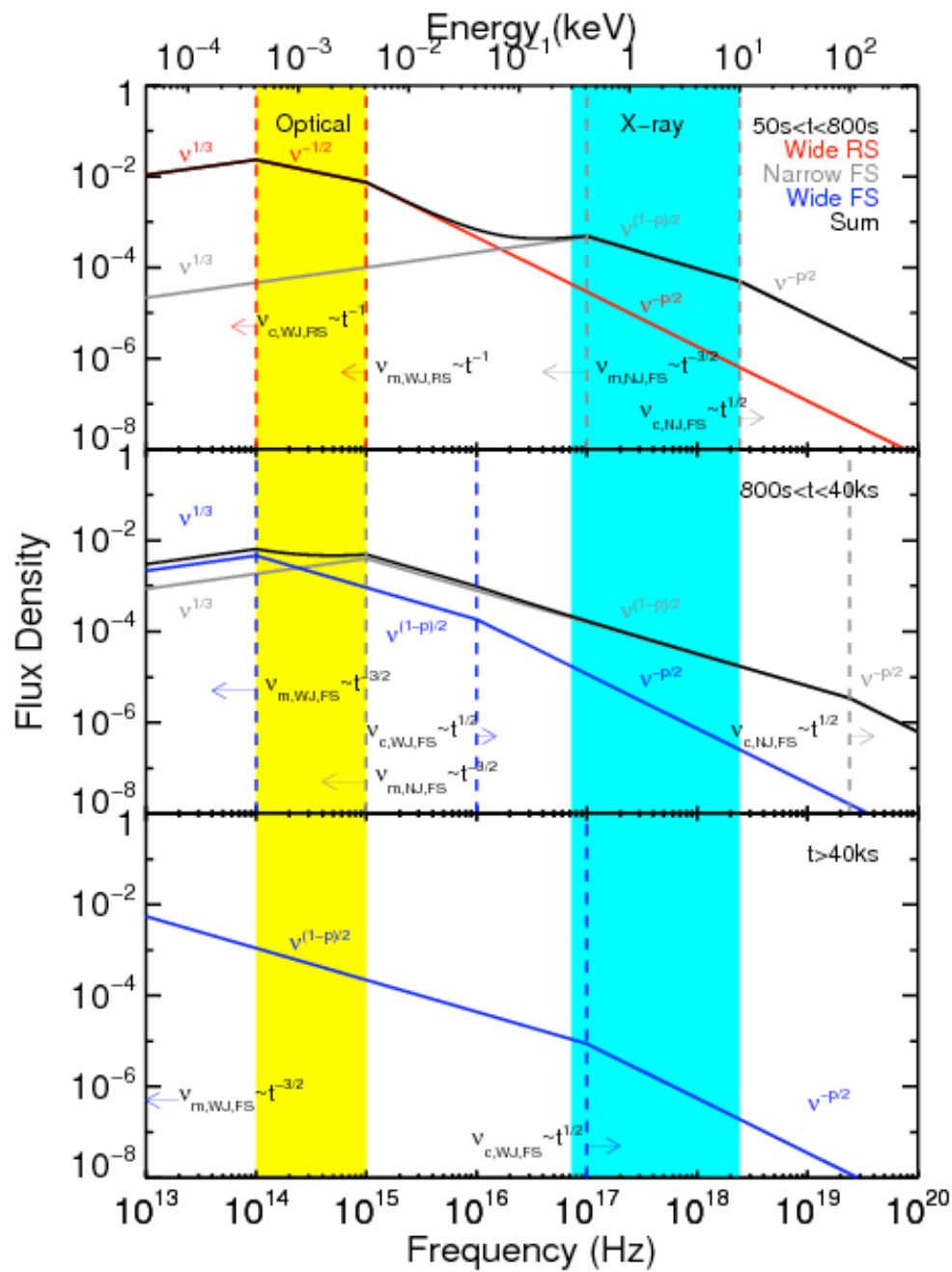

**Supplementary Figure 10 | Spectral Components of Two-Component Jet Model and Dependent Temporal Regimes.** The wide jet reverse shock, narrow jet forward shock, and wide jet forward shock synchrotron spectra and the time dependencies of their frequencies are shown in this schematic. During a portion of the central time period (1-10ks), all three spectral components contribute at some level before the wide jet forward shock dominates, and the reverse shock fades exponentially. The reverse shock is not shown in the central panel because of its wide range of variability, but is implied during the transition between the first and second panel.

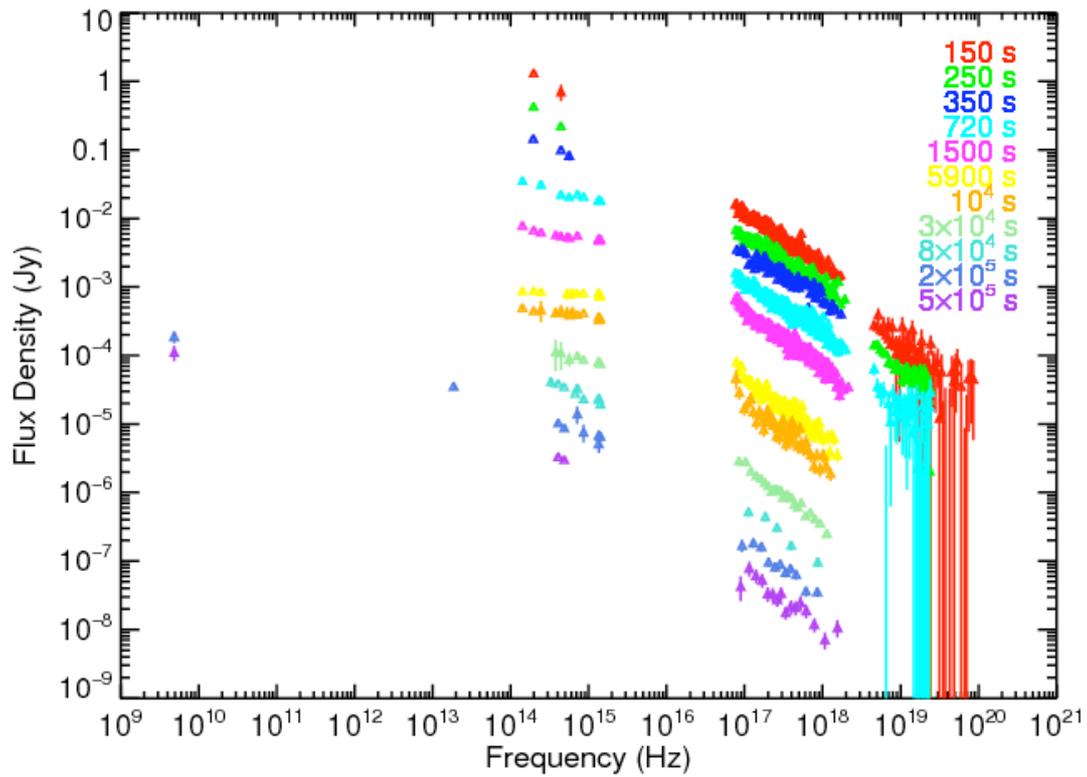

**Supplementary Figure 11 | SEDs with fixed E(B-V)=0.05.** Mean optical extinction is derived from individual fits to each separate SED, not assuming optical and X-ray are necessarily ever on same spectral power-law. Fits to these SEDs with the two-component model are shown in Supplementary Figure 12. Galactic extinction and absorption are taken into account; only intrinsic values at z=0.937 are given above. MW and LMC extinction laws were tested, but SMC[65] consistently provides the best fits (though we note LMC is not ruled out). The SMC law generally best represents the low metallicity dwarf galaxies that host GRBs[66,67]. The Lyman forest correction has been applied to the UVOT data[68]. Solar abundances were adopted for $N_H$ in calculating X-ray absorption.

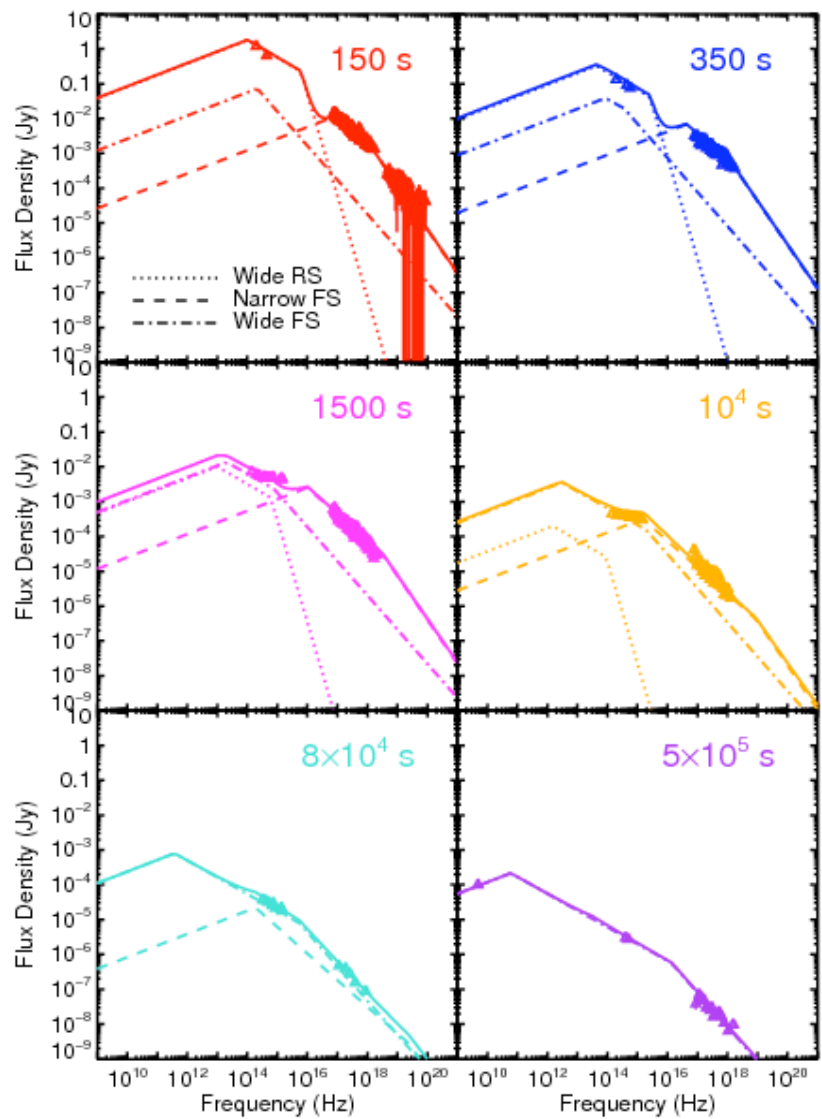

**Supplementary Figure 12 | Two-Component Jet Model SEDs.** Demonstration that evolution of three spectral components evolving like schematic Supplementary Figure 10 can qualitatively reproduce shape of observed SEDs. Normalizations scale like X-ray and optical light curve temporal slopes. Parameters are fine tuned rather than numerically fit due to large range of possible values and starting points. Only every other SED is shown in above plots.

**One-Component Complex Density Medium Model**

This model was motivated by the apparent shape and evolution of the broadband SEDs (Supplementary Figure 13). The basic picture behind this model is that a single spectral component with a cooling break ($\nu_c$) varies, along with differing temporal slopes as the jet probes changes in the density profile. The cooling break begins below the optical band and moves into the spectral regime between the optical and X-ray bands as expected for a decreasing density medium (wind environment). In this model, the optical is decaying faster than the X-rays ($\alpha_{opt}$=2.5 vs. $\alpha_x$=1.45) at early times because the cooling break lies between the bands. At a distance corresponding to the shock location at $T$+1800s in the observed frame, the profile of the medium surrounding the star changes to a constant or increasing density, as could be caused by running into a shell from the progenitor star wind. This causes $\nu_c$ to change direction, heading back to the optical band, which would cause the optical light curve to decay slower than the X-ray ($\alpha_{opt}$=1.25 vs. $\alpha_x$=1.85). The movement of $\nu_c$ must again slow (or stop) at about 40 ks to explain the final temporal slopes in the X-ray and optical both being ~1.2. Another evolution of decreasing density would be required to explain the light curve break at ~1 Ms where the afterglow is too faint to construct a reliable SED.

The observed temporal dependence of $\nu_c$ in this model should translate into a measurement of the density profile of the medium. For $p$<2 ($p$=1.5 from fits), $\nu_c \propto t^{(3k-4)/[2(4-k)]}$, or $k$=(8$x$+4)/(3+2$x$). Here $k$ is the power-law index of the external density profile ($n \propto r^{-k}$), and $x$ is the temporal index of the cooling frequency ($\nu_c \sim t^x$). Using the fitted values of $\nu_c$ from the SEDs, we fit them as a function of time as shown in Supplementary Figure 14. We measure that for $t \lesssim$ 1800 s, $x$=1.08±0.04, and for $t \gtrsim$ 1800 s, $x$=-1.00±0.14. This implies $k(t \lesssim$ 1800 s)=2.5, and $k(t \gtrsim$ 1800 s)=-4.0. The first density profile is not precisely the $r^{-2}$ wind profile, yet it is still physically reasonable. There is substantial evidence supporting wind density profiles steeper than $r^{-2}$ for stellar winds accelerated by radiation pressure from the central star. The late time index would require that the jet has encountered a density enhancement due, perhaps, to a mass ejection event. The existence of this complex medium is observationally supported by the high-resolution VLT/UVES spectra[69]. This complex environment is physically plausible when considering what is observed around local massive stars (such as Eta Carinae).

The details of the spectral evolution are simple in comparison to the two-component model. However, the temporal behaviour is difficult to reconcile, even if one considers all possible alternatives including a very early ($t$ < 100 s) jet break that would allow steeper temporal slopes. We fit both pre- and post-jet break closure relations with and without lateral spreading, and are unable to reproduce the observed temporal behaviour. Perhaps with further modifications to microphysical parameters and model dependencies, one could manufacture a similar model that could adequately reproduce the observed data. Detailed numerical modelling (that is beyond the scope of this paper) is needed to fully explore the consequences of this model.

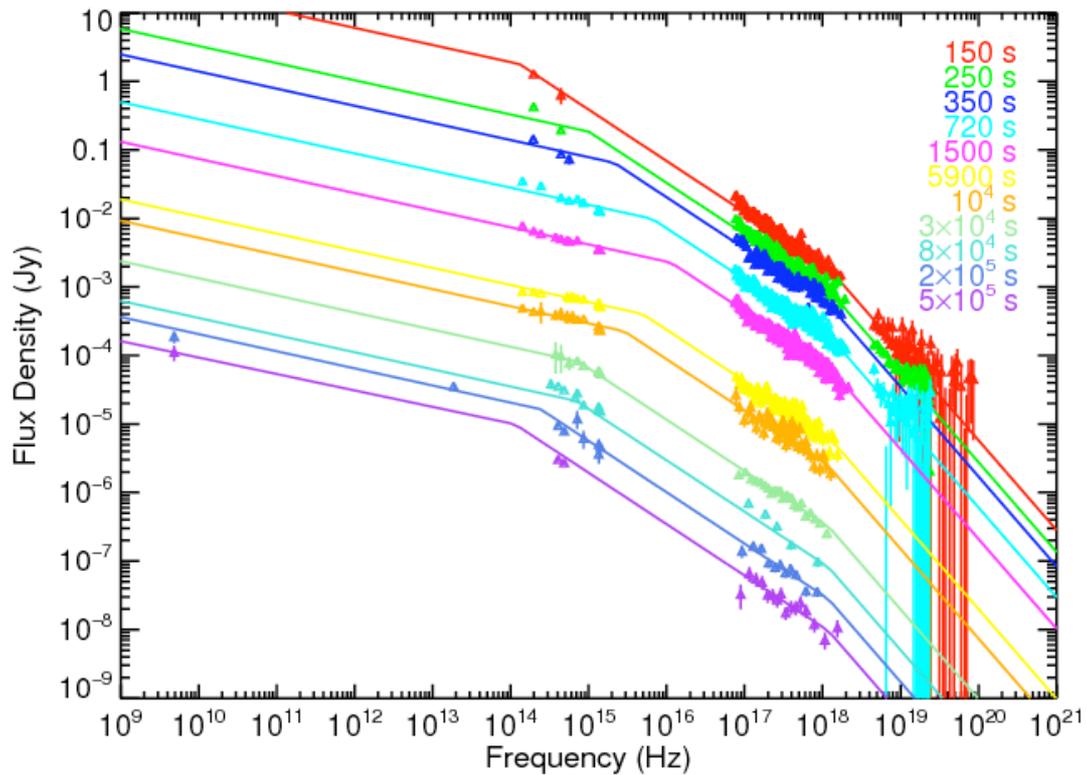

**Supplementary Figure 13 | SED Fits for One-Component Complex Density Structure Model.** Mean optical extinction and X-ray absorption are derived from fits to SEDs at $2\times10^5$ s and $5\times10^5$ s, assuming that the optical and X-ray are on same spectral power-law segment. The model fit of a single spectral component to each SED is shown to demonstrate the evolution of $\nu_c$. The cooling break begins below the optical band, increases into the frequency regime between the optical and X-ray bands and then decreases to below the optical band. These SEDs were constructed using extinction determined by assuming that the last two SEDs have optical and X-ray points on the same spectral segment. The extinction is frozen to E(B-V)=0.028, and X-ray absorption to $N_H=9.2\times10^{20}$ cm$^{-2}$.

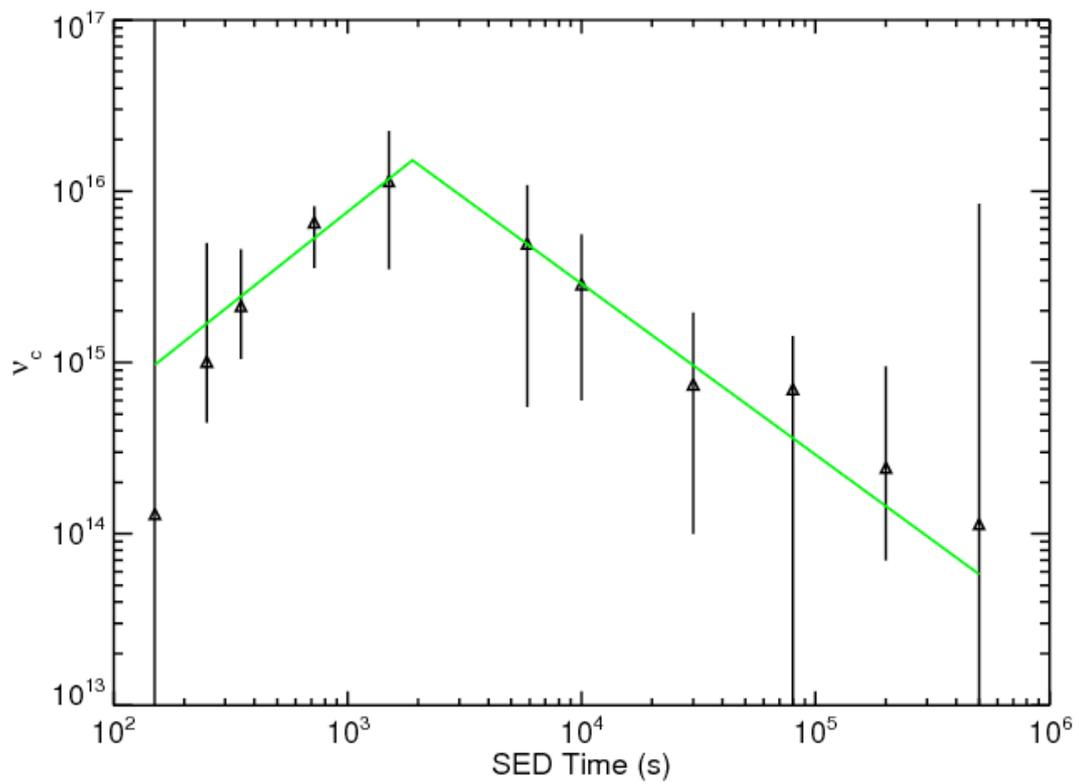

**Supplementary Figure 14 | $\nu_c$ as a Function of Time.** Fits to $\nu_c$ from each SED in Supplementary Figure 13. Best fit temporal dependencies are $t^{+1.08}$ and $t^{-1.00}$, suggesting $n \propto r^{-2.5}$ and $n \propto r^{-4.0}$. The break in the first SED is unconstrained, but is shown as plotted in Figure 13 (just below the optical data) for comparison.

# Supplementary References